\documentclass[a4paper,11pt]{article}

\pdfoutput=1

\usepackage{jheppub}
\usepackage{amsfonts}
\usepackage{graphicx,color}
\usepackage{float}
\usepackage{hyperref}
\usepackage{subfigure}
\usepackage[utf8]{inputenc}
\usepackage[english]{babel}
\usepackage{textcomp}
\usepackage{mathtools}
\usepackage{gensymb}
\usepackage{amsbsy}
\usepackage{capt-of}
\usepackage[T1]{fontenc} 
\usepackage{physics}
\usepackage{multicol} 
\usepackage{multirow}
\usepackage[table]{xcolor}
\usepackage{lscape}
\usepackage[format=plain,small,labelfont={bf},up]{caption} 
\usepackage[title]{appendix}
\usepackage{graphicx}  
\usepackage{subcaption} 
\usepackage{amsmath,amsthm,amssymb}
\usepackage{soul}
\usepackage{blkarray}
\usepackage{array}
\usepackage{booktabs} 

\RequirePackage{color}
\usepackage{colortbl}
\definecolor{denim}{rgb}{0.08, 0.38, 0.74}
\hypersetup{
    colorlinks=true,
    linkbordercolor={white},
    linkcolor={denim},
    citecolor={denim},
    filecolor={denim},      
    urlcolor={denim},
}

\newcommand{\vol}{\mathcal{V}}

\numberwithin{equation}{section}

\def\be{\begin{equation}}
\def\ee{\end{equation}}
\def\bea{\begin{eqnarray}}
\def\eea{\end{eqnarray}}

\def\ie{{\it i.e.~}}

\setcitestyle{square, numbers, comma, sort&compress}

\title{\boldmath Warming up the Fibres}

\author[a]{Dibya Chakraborty}    
\author[b]{and Rudnei O. Ramos }

\affiliation[a]{Centre for Strings, Gravitation and Cosmology, Department of Physics, Indian Institute of
Technology Madras, Chennai 600036, India}
\affiliation[b]{Departamento de F\'{\i}sica Te\'orica,
  Universidade do Estado do Rio de Janeiro,
  20550-013 Rio de Janeiro, RJ, Brazil}

\emailAdd{dibyac@physics.iitm.ac.in}
\emailAdd{rudnei@uerj.br}

\abstract{
Fibre inflationary models are constructed in type-IIB string flux compactification.
These models have been shown to be in agreement with the cosmological observations under appropriate choices of parameters,
which originate from their string theory construction. In the present work, we embed such models, originally studied in 
the cold inflation picture, in the context of warm inflation. We study the viability of different fibre inflation potentials
in both strong and weak dissipative regime of warm inflation. In fibre inflation, the inflaton is a four-dimensional complex 
manifold --- a fibre of a $K3$ fibered Calabi-Yau. The potential in this case is generated by an interplay of various perturbative 
and non-perturbative corrections. The former type of corrections consists of leading-order $\alpha'^3-$ term, 
higher derivative $F^4-$ correction, and various string loop corrections of KK, log-loop and winding type. 
Depending on the balance between several corrections, we present four different fibre inflationary potentials and show 
that the warm inflationary pictures for all of them can successfully fall in the viable window from both Planck and recent Atacama Cosmology Telescope (ACT) data. 
We show that warm inflation makes it possible to extend the range of parameters of applicability of these models.
Our results also indicate that with the help of a large dissipation ratio, one can achieve a sub-Planckian field excursion, 
although, this runs on the possibility of moving away from the perturbative control of low-energy four-dimensional supergravity theory. 
}

\keywords{fibre inflation, warm inflation, observational constraints}

\arxivnumber{ }

\begin{document} 
\maketitle

\section{Introduction}

Although phenomenologically successful, cosmological inflation~\cite{Guth:1980zm,Sato:1981ds,Albrecht:1982wi,Linde:1981mu,Linde:1983gd} still lacks a microphysical understanding within a fundamental theory of quantum gravity, such as string theory. Due to the high-energy scale of inflation, it is largely sensitive to physical effects close to the Planck scale. This fact has motivated the search for inflaton candidates within the realm of string compactifications
\cite{Baumann:2014nda}. String theory features a plethora of scalar fields in 4D effective field theories (EFTs) which has facilitated this search. The scalars can either arise from the closed string sector parametrising the shapes (known as complex-structure (CS) moduli) and sizes (known as K\"ahler moduli (KM)) of the extra-dimensions \cite{Candelas:1985en,Greene:1993vm, Grana:2005jc, Douglas:2006es}, or open string moduli sector parametrising the geometry of the non-perturbative objects such as branes and orientifold planes \cite{Polchinski:1995mt, Grana:2005jc, Douglas:2006es, Marchesano:2007de}.

Once the compactification setup is defined through the internal geometry, one can estimate the higher-order corrections to the inflationary potential. Supersymmetry present in the construction ensures the smallness of such corrections. It has been shown that when higher-order quantum corrections are small enough to generate a flat scalar potential, suitable to study inflation, KM can act as natural inflaton candidates \cite{Conlon:2005jm}. The first step to construct a string theory model is to make sure all the scalar fields arising from the internal geometry are stabilized to their minima and not participate in being the mediator of fifth forces. A large class of string theory model building actually relies on type IIB compactifications \cite{Giddings:2001yu, Kachru:2003aw, Balasubramanian:2005zx, Grana:2005jc, Douglas:2006es} and we shall follow this path as well in this work. 

In the standard paradigm, CS moduli and axio-dilaton are stabilized in the presence of 3-form background fluxes~\cite{Giddings:2001yu}. The low energy type IIB theory can be approximated to be the supergravity effective theory, in which the effective Lagrangian can be written in terms of a holomorphic superpotential and a real K\"ahler potential. The presence of fluxes in type IIB are manifested in the tree-level superpotential~\cite{Gukov:1999ya}. By construction,  this superpotential depends only on the CS moduli and the axio-dilaton. On the other hand, although the K\"ahler potential depends on all the moduli, but they form a block-diagonal metric. Due to these nature, the tree-level F-term scalar potential depends on CS and axio-dilaton but not on KM. This is at the heart of the well-known \emph{no-scale structure} of the theory. Minimizing the F-term scalar potential stabilizes the CS and the axio-dilaton, leaving the theory with a constant superpotential term $W_0$ at the minimum.
To remedy this situation, that is to address the stabilization of the KM, one has to break the no-scale structure by incorporating corrections beyond tree-level \'a la quantum corrections. These are known as next to leading order effects or higher order terms in perturbative or non-perturbative expansions in type IIB theory. They can be non-perturbative corrections to the superpotential~\cite{Kachru:2003aw,Witten:1996bn,Balasubramanian:2005zx}, string loops~\cite{vonGersdorff:2005bf,Berg:2004ek,Berg:2005ja,Berg:2007wt,Cicoli:2007xp,Cicoli:2008va,Antoniadis:2018hqy,Antoniadis:2019rkh,Gao:2022uop}, or higher derivative corrections~\cite{Becker:2002nn,Ciupke:2015msa}. These corrections lift the flat KM and stabilize the whole system to anti-de Sitter (AdS) vacua. Next step, for phenomenological purposes, is to add a supersymmetry breaking \emph{uplift} contributions to the scalar potential. It can either be a D-term uplift~\cite{Burgess:2003ic}, 
a T-brane uplift~\cite{Cicoli:2015ylx} or an anti-brane uplift~\cite{Kachru:2002sk,Crino:2020qwk,Vafa:2005ui,Cicoli:2024bxw}. {}Finally, if the scalar potential is shallow enough along any of the direction(s) of the KM, it can drive an early period of accelerated expansion --- inflation --- before reaching their respective minimum of the potential. These models are globally classified as KM inflationary models (see ref.~\cite{Cicoli:2023opf} and the references therein). 

A promising class of such KM inflation is Fibre Inflation (FI)~\cite{Cicoli:2008gp}, where the inflaton is the K\"ahler modulus, which can be parametrised as either the fibre modulus (4-cycle) itself or a ratio of two large 4-cycles of a K3 fibered Calabi Yau (CY) manifold --- orthogonal to the overall volume. They act as an ideal inflaton candidate since it is the leading-order flat direction that enjoys an approximate non-compact shift symmetry~\cite{Burgess:2014tja,Burgess:2016owb}. Depending on the nature of quantum corrections to be added, different versions of fibre inflation have been proposed over the years: the original model~\cite{Cicoli:2008gp} is based on winding and KK string loops, while in the ref.~\cite{Cicoli:2016chb} the authors have exploited the effects of KK loops and $F^4-$corrections. {}Finally in the ref.~\cite{Broy:2015zba}, effects of winding loops and $F^4$ have been considered. Besides, extending the above studies with the addition of global embedding have also been studied extensively in the refs.~\cite{Cicoli:2016xae,Cicoli:2017axo,AbdusSalam:2022krp,Bera:2024ihl}. FI models have received considerable attention due to their theoretical robustness, as well as their resemblance to Starobinsky inflation~\cite{Broy:2014xwa,Brinkmann:2023eph}. 

Given the above motivations, possible realization of warm inflationary dynamics in these class of models might act as an important achievement. Warm inflation (WI)~\cite{Berera:1995ie}, proposed almost two decades ago --- is an alternative inflationary scenario where as inflation progresses the inflaton dissipates its energy to a nearly constant radiation bath. As a result, at the end of inflation, the Universe does not need to go through a reheating phase because the Universe does not remain cold any longer in the previous phase (see refs.~\cite{Kamali:2023lzq,Berera:2023liv} for recent reviews on WI). Another encouraging fact about WI is that it aligns with effective field theory (EFT) and has the potential to find an ultraviolet (UV) completion in quantum gravity~\cite{Das:2018rpg,Motaharfar:2018zyb,Das:2019acf,Berera:2019zdd,Kamali:2019xnt,Berera:2020iyn,Das:2020xmh,Brandenberger:2020oav,Motaharfar:2021egj}. Hence, constructing WI models in a concrete quantum gravity framework has become increasingly crucial.
It should also be noticed that WI has possible signatures that make it 
distinguishable from cold inflation. These include, for example, differences in non-Gaussianities~\cite{Bastero-Gil:2014raa}, 
differences in the  consistency relation between the tensor-to-scalar ratio and the tilt of the tensor spectrum, with  WI in general having  $r< 8|n_t|$ (see, e.g. refs.~\cite{Bartrum:2013fia,Benetti:2016jhf}),
and also the possibility of having matter isocurvature perturbations (originating from the dissipative effects during WI),
which are fully anti-correlated with the dominant adiabatic curvature perturbations~\cite{Bastero-Gil:2014oga}. 

In WI, the presence of the thermal bath can potentially lead to temperature-dependent correction terms. These terms are known to disrupt and prevent WI 
to happen~\cite{Berera:1998gx,Yokoyama:1998ju,Moss:2008yb,delCampo:2010by,BasteroGil:2012zr}. Thus, models able to explicitly realize WI
have used specific symmetries to protect the inflaton potential to receive large (quantum or thermal) corrections, like making use
of supersymmetry (SUSY)~\cite{Berera:2008ar,BasteroGil:2012cm} or symmetries satisfied by pseuso-Goldstone bosons (making the role of the inflaton), like
in refs.~\cite{Bastero-Gil:2016qru,Bastero-Gil:2019gao,Berghaus:2019whh}.
In the string motivated models that we study here, the inflaton field is a saxion --- the real part of a chiral multiplet. 
In this context, as already stated in e.g.~ref.~\cite{Brinkmann:2023eph}, and further elaborated in the refs.~\cite{Burgess:2014tja,Burgess:2016owb}, 
in the presence of a saxion, especially the K\"ahler moduli that we consider, if it represents a field orthogonal to the overall volume, like the fibre modulus for instance, then the field can also enjoy an effective and approximate non-compact shift symmetry.
This approximate non-compact shift symmetry can act to protect the inflaton potential from dangerous thermal correction, much similar to what happens in axion-like constructions of WI~\cite{Berghaus:2019whh,Laine:2021ego,Berghaus:2025dqi}. This particular type of shift symmetry is a consequence of the fact that to leading order, the potential depends only on the volume and, therefore, in the base-fibre space, there is only one flat direction which is along the fibre modulus. Now, flat direction leads to shift symmetry, hence the shift symmetry of the fibre remains intact. Next, in order to lift the flat direction, next-to-leading order corrections in volume are usually added, which again results in a cancellation due to extended no scale structure~\cite{Cicoli:2007xp}. In the end, the potential for the fibre modulus becomes sufficiently suppressed, and hence an effective shift symmetry takes place. This mechanism applies to both quantum and thermal corrections to the potential.\par

The scaling shift symmetry also manifests itself in the coupling of inflaton to hidden/visible sector fields. In particular, axions are a generic source of hidden sector fields. In WI,  inflaton or the fibre modulus decay to these axions, producing extra neutrino-like species. However, the inflaton-axion-axion and inflaton-Higgs interaction terms\footnote{The Langrangian describing this inflaton-axion-axion $(\hat{\phi}\theta_1\theta_2)$ interactions can be written in terms of their kinetic mixing as (see, for intance ref.~\cite{Cicoli:2018cgu}):
    \begin{equation*}
        \mathcal{L}_{\hat{\phi}\theta_1\theta_2}=\frac{1}{\sqrt{3}}\frac{m_{\phi}^2}{M_{pl}}\hat{\phi}\frac{\theta_1^2}{2\langle\tau_1\rangle^2}-\frac{1}{2\sqrt{3}}\frac{m_{\phi}^2}{M_{pl}}\hat{\phi}\frac{\theta_1^2}{\langle\tau_2\rangle^2},
    \end{equation*}
    where $\theta_1$ is the axion of the fibre and $\theta_2$ is the axion of the base and $m_{\phi}^2\simeq \frac{m_{3/2}^2}{\mathcal{V}^{4/3}}\sim \frac{W_0^2}{\mathcal{V}^{10/3}}$. Similarly, the coupling term of inflaton to the visible sector (e.g. to the Higgs boson) arises through the Giudice-Masiero term and can be written as:
    \begin{equation*}
        \mathcal{L}_{GM}=-\frac{z}{8\sqrt{3}}\frac{m_{\phi}^2}{M_{pl}}\hat{\phi}\left(\hat{H_u}\hat{H_d}+h.c.\right),
    \end{equation*}
    where $z$ being an order one constant, makes the interaction term again suppressed in volume due to the presence of $m_{\phi}^2$.} as shown in \cite{Cicoli:2018cgu} are suppressed by the inflaton mass, or in other words, are volume suppressed. Even though these interactions will generate suppressed thermal and quantum corrections to the inflaton potential, they can still contribute to the radiation component.  \par

The possible embedding of WI in string models, like the one considered in the present paper, hence, 
can have a similar type of interactions as in the earlier SUSY WI models and more recent studies. This then motivates us to analyze two of the most 
used forms of dissipation coefficients considered in the literature: a dissipation coefficient that is linear in the temperature (see, e.g. ref.~\cite{Bastero-Gil:2016qru}) and one that has a cubic dependence on the temperature (see, e.g., refs.~\cite{Berera:2008ar,BasteroGil:2012cm,Berghaus:2019whh})\footnote{For a recent review on different model realizations of WI, see ref.~\cite{Kamali:2023lzq} and also references therein.}.

Throughout this paper, we work with natural units, where the speed of light, Planck's constant, and Boltzmann's constant are all set to $1$, 
$c=\hbar=k_B=1$. We also adapt our convention to the reduced Planck mass, defined as $M_{\rm Pl}=(8\pi G)^{-1/2}\simeq 2.44\times 10^{18}\,\mathrm{GeV}$, where $G$ is Newton's gravitational constant. 

The paper is organized as follows. In section~\ref{sec:2}, we briefly review the notable features of FI, putting special emphasis on the quantum corrections. Depending on the nature of the sub-leading corrections, we present four different models of FI developed in the literature over the years in section~\ref{section_fourFibre}. These will be the four type of fibre inflation models that we will analyze. In section~\ref{sec:3}, we present the generic features of any WI model. In section~\ref{sec:numerical_results}, we present the successful embedding of FI in WI framework and present our results, contrasting them with the observations constraints. In section~\ref{sec:conclusion} we give our conclusions and final remarks.

\section{A brief review of fibre inflation}\label{sec:2}

We start our discussion by laying out the basics of FI. To do so, we first demonstrate two of the main heavy KM stabilization scenarios: large volume scenario (LVS) and perturbative LVS (pLVS). In addition, we also emphasize the possible quantum string corrections to stabilize the lighter KM --- leading-order flat direction. 

\subsection{Contributions to the scalar potential}

The general formula to calculate the F-term scalar potential from type IIB superstring compactifications is encoded in:
\be
V=e^K\left(K^{A\overline{B}}\,D_AW \,D_{\overline{B}}\overline{W}-3|W|^2\right),\qquad D_AW\equiv W_A+K_AW,
\ee
and which governs the low-energy dynamics of the four-dimensional $\mathcal{N}=1$ effective supergravity theory. In the above equation, 
$K$ denotes the K\"ahler potential and $W$ is the holomorphic superpotential. Both are functions of the CS moduli $z^i$, the axio-dilaton $(S)$, and the KM $T_{\alpha}$. The tree-level K\"ahler metric is diagonal and takes the form:
\be
K_0=K_{CS}-\ln\,\left[-i\,(S-\overline{S})\right]-2\,\ln \,Y,
\ee
where $K_{CS}$ is the CS moduli dependent part, while $Y$ generically depends on the volume/dilaton pieces, such that at tree level, $Y=\mathcal{V}$, where $\mathcal{V}=\frac{1}{3!}k_{\alpha\beta\gamma}t^{\alpha}t^{\beta}t^{\gamma}$ is the volume of the underlying CY manifold. $t^{\alpha}$'s are the two-cycle volumes and $k_{\alpha\beta\gamma}$ is the triple intersection number of the CY-3-fold (CY3). On the other hand, the superpotential is only a function of CS and axio-dilaton and depends on the KM modulo the addition of non-perturbative (NP) effects, to be discussed soon. The holomorphic superpotential is:
\be
W=W_{\text{flux}}+W_{np}(S,T_{\alpha}),
\ee
where 
\begin{equation*}
    W_0=\left\langle W_{\text{flux}}\right\rangle,
\end{equation*}
and $W_{\text{flux}}$ is the superpotential induced by the 3-form fluxes $(F_3,H_3)$ \cite{Gukov:1999ya}. Due to the diagonal decoupled structure of 
$K\equiv K_{CS}+K(S,T_{\alpha})$, the inverse metric takes a block-diagonal form, producing a scalar potential:
\begin{align*}
    &V\equiv V_{CS}+V_K,\,\qquad V_{CS}=e^K K^{i\overline{j}}_{CS}\,D_iW\,D_{\overline{j}}\overline{W},\\
   & V_K=e^K\left(K^{A\overline{B}}\,D_AW \,D_{\overline{B}}\overline{W}-3|W|^2\right),\,\,\,A,B\in \left\{S,T_{\alpha}\right\}.
\end{align*}
Moduli stabilization in low-energy type IIB flux-compactification is a two-step procedure. In the first step, CS and axio-dilaton are stabilized by the supersymmetric flatness condition $D_{U^i}W_{\text{flux}}=0=D_SW_{\text{flux}}$. After the first step, one obtains $W_0$, leaving the KM as a flat direction. Performing the summation in $V_K$ in $\{S,T_{\alpha}\}$, one can show that $V_K$ becomes zero at tree level --- this is the so-called \emph{no-scale structure} of the scalar potential. As a second step, to lift the flat direction along the KM, one has to incorporate sub-leading corrections such as either perturbative corrections in inverse-string-tension $\alpha'\,(=2\pi l_s^2)$ and string loop $(g_s)$ or non-perturbative corrections. 

We shall now list out a possible set of corrections --- susceptible to lift the flat directions along the KM~\cite{Cicoli:2007xp,Cicoli:2021rub,AbdusSalam:2020ywo,Antoniadis:2018hqy}:

\begin{itemize}

    \item \textbf{$\alpha'-$ corrections:} the first corrections to be considered is $\mathcal{O}(\alpha'^3)$ corrections to the tree-level K\"ahler potential by changing the volume~\cite{Becker:2002nn}:
    \begin{equation}
        Y=\mathcal{V}+\frac{\hat{\xi}}{2},\qquad \hat{\xi}=\frac{\xi}{g_s^{3/2}}=-\frac{\zeta(3)\chi}{2(2\pi)^3},
    \end{equation}
    where $\chi$ is the Euler characteristics of a CY. This gives rise to a potential of the form:
    \be\label{alpha-corr}
    V_{\alpha'}=g_s\frac{3\hat{\xi}|W_0|^2}{4\mathcal{V}^3}.
    \ee

    \item \textbf{$g_s$ loop corrections $K_{g_s}^{KK}$ and $K_{g_s}^W$:} These corrections arise at the level of $\mathcal{O}(g_s^2\alpha'^4)$ and $\mathcal{O}(g_s^2\alpha'^2)$ correcting the K\"ahler potential to~\cite{Berg:2004ek,Berg:2005ja,Berg:2007wt}:
\begin{equation}
        K_{g_s}^{KK}=g_s\sum_{\alpha} \frac{c_{\alpha}^{KK}t_{\alpha}^{\perp}}{\vol},\qquad K_{g_s}^W=g_s\sum_{\beta}\frac{c_{\beta}^w}{t^{\cap}_{\beta}\vol},
\end{equation}
where $t_i^{\perp}$ denotes the 2-cycle volume moduli perpendicular to the parallel branes and $t^{\cap}_j$  is the intersection 2-cycle volume moduli of the $j-$th couple of intersecting branes. They correct the scalar potential to:
    \begin{align}\label{corr_kk}
        & V_{g_s}^{KK}=\frac{g_s^3}{2}\frac{|W_0|^2}{4\vol^2}\sum_{\alpha\beta}c_{\alpha}^{KK}c_{\beta}^{KK}(2t^{\alpha}t^{\beta}-4k^{\alpha\beta}),\\
        & V_{g_s}^w=-\frac{g_s |W_0|^2}{\vol^2}\sum_{\alpha}\frac{c_{\alpha}^w}{t^{\cap}_{\alpha}\vol}, \label{corr_w}
    \end{align}
where $k^{\alpha\beta}=\frac{\partial t^{\alpha}}{\partial \tau_{\beta}}$, with $\tau_{\beta}=\partial_{t^{\beta}}\vol$ being the 4-cycle volume moduli. Besides, one finds that roughly $c_{\alpha}^{KK}\simeq c_{\alpha}^w\simeq \frac{1}{128\pi^4}$.
    
    \item \textbf{Higher-derivative $\alpha'$ corrections to the supergravity action:} These represents $\mathcal{O}(\alpha'^3)$ corrections, 
which are reflected in the 4D EFT as an $F^4$ correction of the form~\cite{Ciupke:2015msa}:
\be
    V_{F^4}=-\frac{\lambda}{4}g_s^{1/2}\frac{|W_0|^4}{\vol^4}\Pi_{\alpha}t^{\alpha},\label{corr_F4}
\ee
where $\Pi_{\alpha}$ are the second Chern numbers of the CY3 and $\lambda$ is a dimensionless constant of the order $\mathcal{O}(10^{-2})-\mathcal{O}(10^{-4})$ \cite{Grimm:2017okk}.
    
    \item \textbf{Log-loop corrections:} Their origin can be traced back to the $R^4$ term appearing in the effective type IIB sugra action --- correcting the K\"ahler potential to~\cite{Antoniadis:2018hqy, Antoniadis:2019rkh}:
\be
    Y=\vol+\hat{\eta}(\ln\vol-1),\quad \hat{\eta}=-g_s^2\hat{\xi}\frac{\zeta[2]}{\zeta[3]},\quad \hat{\xi}=\frac{\xi}{g_s^{3/2}}.
\ee
This new addition adds a correction to the scalar potential of the form:
\be
    V_{log-loop}=\frac{3g_s\hat{\eta}}{16\pi\vol^3}(\ln\vol-2)|W_0^2|.
\ee

    \item \textbf{Non-perturbative effects:} Non-perturbative (NP) effects are added at the superpotential which read~\cite{Witten:1996bn}:
\be\label{np-superpotential}
     W_{np}=A_se^{-a_sT_s},
\ee
where, for convention,  $T_s=\tau_s+i\,\theta_s$ is the blow-up modulus, while for Euclidean D3-brane $a_s=2\pi$.

    \item \textbf{dS uplifting contribution:} The uplift contribution is mainly originated from D3-branes, T-branes or D-terms with an effective term equal to:
\be
    V_{up}=\frac{C_{up}}{\vol^{n/3}},\qquad n<9,\qquad C_{up}>0.
\ee

\end{itemize}

Note that $W_0$ does not depend on the K\"ahler moduli and, therefore, can be considered to be constant after stabilizing the CS moduli and axio-dilaton. Next, we aim to outline two significant heavy moduli stabilization schemes in which the inclusion of certain leading quantum corrections, as discussed above, helps stabilize some of the KM present in the compactification, while leaving a flat direction that will be further lifted by sub-leading corrections.

\subsection{Large volume scenario stabilization scheme}

Fibre inflationary models rely on a specific form of CY, namely a K3 or $\mathbb{T}^4$ fibration over a $\mathbb{P}^1$ base. The overall volume of the CY can be expressed as:
\begin{align}\label{fibre_vol}
    \mathcal{V}=\alpha(\sqrt{\tau_1}\tau_2-\gamma \tau_3^{3/2}),
\end{align}
where $\tau_1$ corresponds to the volume of a $T^4$ or $K3$ fibered over a $\mathbb{P}^1$ base. $\tau_2$ is the geometric modulus corresponding to the volume of the base. $\tau_3$ is the blow-up modulus whose size determines the strength of the non-perturbative effects added to the superpotential. $\alpha$ and $\gamma$ are order one constants determined by the intersection numbers of the four cycles. The LVS scenario utilizes the interplay of non-perturbative effects \eqref{np-superpotential} and leading $\mathcal{O}(\alpha'^3)$ correction \eqref{alpha-corr} and work in the approximation when $\sqrt{\tau_1}\tau_2>\tau_3>1$. In this limit, the LVS scalar potential, without the uplift contribution, reads \cite{Balasubramanian:2005zx} \cite{Cicoli:2007xp}:
\be
V_{LVS}\simeq g_s\frac{8a_s^2A_s^2\sqrt{\tau_3}e^{-2a_s\tau_3}}{3\vol}+4g_s|W_0|a_sA_s\cos(a_s\theta_3)\frac{\tau_3e^{-a_s\tau_3}}{\vol^2}+\frac{3\xi|W_0|^2}{4\sqrt{g_s}\vol^3}.
\ee
Minimizing the scalar potential gives the result:
\be
\langle\tau_3\rangle\sim \xi^{2/3},\qquad \langle\vol\rangle\sim \frac{W_0\sqrt{\langle\tau_3\rangle}}{a_sA_s}e^{a_s\langle\tau_3\rangle},\qquad \langle\theta_3\rangle=\frac{\pi}{a_s}(1+2k_s),\,\,k_s\in \mathbb{Z}.
\ee
At this stage, $\tau_3$ and the combination $\vol\simeq \sqrt{\tau_1}\tau_2$ are stabilized. However, $\tau_1,\; \tau_2$, or a combination of them orthogonal to $\vol$, are not stabilized. Hence, the flat direction can be lifted by adding either loop corrections of winding type and KK-type and higher derivative corrections or log-loop corrections. In this paper, we discuss four different FI models based on their addition of sub-leading corrections. The original FI~\cite{Cicoli:2008gp}, which we call \emph{Type-I FI}, uses string loop correction of winding and KK type. \emph{Type-II FI}~\cite{Cicoli:2016chb} uses winding loops and $F^4$ correction. \emph{Type-III FI} \cite{Cicoli:2016xae} utilizes loop corrections (both KK and winding type) and $F^4-$terms. {}Finally, \emph{type-IV FI} \cite{Bera:2024ihl} uses log-loop correction to be the heavy moduli stabilization scheme and utilizes winding loops and higher derivative corrections to stabilize the leading order flat direction. 

{}For the sake of brevity, when one considers the volume form \eqref{fibre_vol}, the correction terms introduced above (\eqref{corr_kk},\eqref{corr_w}, and \eqref{corr_F4}) take the following simplified closed forms in terms of the fibre modulus $\tau_1$ and the leading order stabilized volume $\vol$:
\begin{table}[H]
\begin{center}
\centering
    \resizebox{0.9\textwidth}{!}{ 
    \begin{tabular}{| l | c | c |}
\hline
\cellcolor[gray]{0.9} $V_{g_s}^{KK}$ &\cellcolor[gray]{0.9} $V_{g_s}^w$ &  \cellcolor[gray]{0.9} $V_{F^4}$  \\
\hline \hline
 $\frac{|W_0|^2}{\langle\vol\rangle^2}g_s^3\frac{(c_1^{KK})^2}{\tau_1^2}$ + $\frac{|W_0|^2}{\langle\vol\rangle^2}g_s^3\frac{2(\alpha c_2^{KK})^2\tau_1}{\langle\vol\rangle^2}$ &  $-\frac{|W_0|^2}{\langle\vol\rangle^2}\frac{g_s}{8\pi}\frac{c^w}{\langle\vol\rangle\sqrt{\tau_1}}$ & $-\left(\frac{g_s}{8\pi}\right)^2\frac{\lambda W_0^4}{g_s^{3/2}\langle\vol\rangle^4}\left(\Pi_1\frac{\langle\vol\rangle}{\tau_1}+\Pi_2\sqrt{\tau_1}\right)$\\
\hline
\end{tabular}}
\end{center} 
\end{table}

\subsection{Perturbative large volume scenario stabilization scheme}\label{subsec:pLVS}

The fundamental difference between LVS and this type of heavy moduli stabilization mechanism is the fact that the latter ignores NP-effects altogether. pLVS uses leading $\alpha'-$ correction and log-loop effects:
\be
V_{pLVS}=\frac{3g_s\hat{\xi}}{32\pi \vol^3}|W_0|^2+\frac{3g_s\hat{\eta}}{16\pi\vol^3}(\ln\,\vol-2)|W_0|^2.
\ee
This subsequently results in an exponentially large volume equal to:
\be
\langle\vol\rangle\sim e^{\frac{a}{g_s^2}+b},\qquad a=\frac{\zeta[3]}{2\zeta[2]}\simeq 0.37,\qquad b=\frac{7}{3}.
\ee
In the following section, we provide a detailed discussion of the four types of FI utilized in our study.

\section{Four types of fibre inflation}\label{section_fourFibre}

More explicitly, the four types of FI that we implement in the context of WI are the following.

\subsection{Type-I fibre}

The fibre model was originally proposed with the aim of lifting the flat $\tau_1$ direction after LVS type moduli stabilization. The addition of loop corrections of winding and KK type are added giving the corrected potential of the form \cite{Cicoli:2008gp}:
\be\label{vsubI}
V_{sub}^{I}(\tau_1)=\frac{|W_0|^2}{\langle\vol\rangle^2}\left[g_s^3\frac{(c_1^{KK})^2}{\tau_1^2} + g_s^3\frac{2(\alpha c_2^{KK})^2\tau_1}{\langle\vol\rangle^2} -\frac{g_s}{8\pi}\frac{c^w}{\langle\vol\rangle\sqrt{\tau_1}}\right].
\ee
After incorporating the subleading corrections, one makes sure that the minimum is a Minkowski one. At the same time, a canonical normalization of $\tau_1=\langle \tau_{1}\rangle e^{k\varphi'}$ with $k=2/\sqrt{3}$ is performed by shifting the zero of the field to its minimum with $\varphi'=\langle \varphi'\rangle+\varphi$ where $\varphi=0$ implies $\varphi'$ modulus is at its minimum. This shift in the canonical scalar field leads us to write the potential in terms of the canonically normalized field $\varphi$ as:
\begin{align}
V_I(\varphi)&=\langle V_{LVS}\rangle+\langle V_{up}\rangle+V_{sub}(\varphi),\nonumber\\
&=V_0\left[3+\delta-4e^{-\varphi/\sqrt{3}}+e^{-4\varphi/\sqrt{3}}+R (e^{2\varphi/\sqrt{3}}-1)\right],\label{typeIpot}
\end{align}
where
\be\label{params1}
R\propto g_s^4\left(\frac{c_1^{KK}c_2^{KK}}{c^w}\right)^2\ll 1,\qquad V_0\propto \frac{C_1}{\langle\vol\rangle^{10/3}},\qquad C_1=\frac{W_0^2(\alpha c_w)^{4/3}}{(g_s c_1^{KK})^{2/3}}.
\ee
The parameter $\delta$ accounts for the additional adjustment of the uplift that might be necessary to maintain $V(\langle\varphi\rangle)=0$, depending on the values of the model parameters, such as $R$ in this case.

\subsection{Type-II fibre}

Type-II fibre~\cite{Cicoli:2016chb} uses winding loop and $F^4-$corrections after stabilizing the volume and the blow-up by the LVS 
mechanism, giving rise to the following potential:
\be
V_{sub}(\tau_1)=\left(\frac{g_s}{8\pi}\right)^2\frac{|\lambda| W_0^4}{g_s^{3/2}\langle\vol\rangle^4}\left(\Pi_1\frac{\langle\vol\rangle}{\tau_1}+\Pi_2\sqrt{\tau_1}\right)-\frac{|W_0|^2}{\langle\vol\rangle^2}\left( \frac{g_s}{8\pi}\frac{c^w}{\langle\vol\rangle\sqrt{\tau_1}}\right).
\ee
Upon canonical normalization and considering the leading term in $F^4-$correction, $V_{sub}$ can be recasted as:
\begin{align}
V_{II}(\varphi)&=\langle V_{LVS}\rangle+\langle V_{up}\rangle+V_{sub}(\varphi),\nonumber\\
&=V_0\left[1+\delta+e^{-2\varphi/\sqrt{3}}-2e^{-\varphi/\sqrt{3}}+R(e^{\varphi/\sqrt{3}}-1)\right],\label{typeIIpot}
\end{align}
where
\begin{align}\label{params2}
& V_0=\frac{3}{2}\left(\frac{g_s}{8\pi}\right)\frac{W_0^2}{\langle\vol\rangle^3}\frac{2C_1}{3\langle\tau_1\rangle},\qquad R=\frac{\Pi_2}{\Pi_1}\frac{\langle\tau_1\rangle^{3/2}}{\langle\vol\rangle},\qquad C_1=\left(\frac{g_s}{8\pi}\right)\frac{|\lambda|W_0^2\Pi_1}{g_s^{3/2}},\nonumber\\
&  C_2=\left(\frac{g_s}{8\pi}\right)\frac{|\lambda|W_0^2\Pi_2}{g_s^{3/2}},\qquad \langle\tau_1\rangle=\frac{4C_1^2}{B^2},\qquad B=4\alpha c^w,
\end{align}
and $\delta$ has the similar meaning as in the type-I fibre potential eq.~(\ref{typeIpot}).

\subsection{Type-III fibre}

Type III fibre inflation~\cite{Cicoli:2016xae} utilizes LVS type moduli stabilization for volume and blow-up modulus and considers loop correction of KK and winding type and $F^4-$corrections. The loop-corrected part will resemble the potential~\eqref{vsubI}, but depending on the topological numbers of the CY ($\Pi_i$ s), the higher derivative terms are re-recast in the form:
\be\label{vsubIII}
V_{sub}^{III}(\tau_1)=V_{sub}^I(\tau_1)-\left(\frac{g_s}{8\pi}\right)^2\frac{\lambda W_0^4}{g_s^{3/2}\langle\vol\rangle^4}\left(24\frac{\langle\vol\rangle}{\tau_1}+36\sqrt{\tau_1}\right).
\ee
Upon canonical normalization, $V_{sub}$ can be re-cast as:
\begin{align}
V_{III}(\varphi)&=\langle V_{LVS}\rangle+\langle V_{up}\rangle+V_{sub}(\varphi),\nonumber\\
&=V_0\left(E+\delta+e^{-4\varphi/\sqrt{3}}-4e^{-\varphi/\sqrt{3}}+R_1e^{2\varphi/\sqrt{3}}+R_2e^{\varphi/\sqrt{3}} \right),\label{typeIIIpot}
\end{align}
where $\delta$ here again has a similar meaning as in the type-I model and the rest of the parameters of this model can be expressed as:
\begin{align}\label{params3}
& V_0=\frac{A|W_0|^2}{\langle\vol\rangle^2\langle\tau_1\rangle^2},\qquad \langle\tau_1\rangle=\left(\frac{4}{C_1}\right)^{2/3}\vol^{2/3},\qquad C_1=\left(\frac{2}{C_1^{KK}}\right)^2\frac{C_w}{g_s^2},\nonumber\\
& A=\frac{g_s}{32\pi}g_s^2(C_1^{KK})^2,\qquad R_1=\left(\frac{C_1^{KK}C_2^{KK}}{C_w}\right)^2\frac{g_s^4}{18},\qquad R_2=\frac{18W_0^2}{\pi}\frac{(C_1^{KK})^{4/3}}{C_w^{5/3}}\frac{|\lambda|g_s^{5/6}}{\langle\vol\rangle^{1/3}},\nonumber\\
&\langle V_{LVS}\rangle+\langle V_{up}\rangle=V_0E=V_0(3-R_1-R_2).
\end{align}

\subsection{Type-IV fibre}

In contrast to the first three previous instances, the type IV fibre uses the pLVS mechanism to stabilize the heavy KMs as shown in subsection~\ref{subsec:pLVS}. In addition, the volume form of eq.~\eqref{fibre_vol} takes a different form: $\vol=2t^1t^2t^3=\frac{1}{\sqrt{2}}\sqrt{\tau_1\tau_2\tau_3}$. One can readily notice that there are no blow-up moduli present in the pLVS setup. Hence, a large volume can be easier to achieve in this scenario. The absence of an exceptional del-Pezzo divisor prohibits us from turning on any NP effects. The type IV FI model was studied in ref.~\cite{Bera:2024ihl}, where the leading order flat direction was lifted by incorporating the leading order $\alpha'$ correction, log-loop correction, winding-type string loop correction and higher derivative $F^4-$corrections. After canonical normalization, the scalar potential takes the following form:
\be
\label{typeIV_pot}
V_{IV}(\varphi) = V_0 \left(C_{up}+\delta + R_0 e^{-2\varphi/\sqrt{3}} - e^{-\varphi/\sqrt{3}}+R_1 e^{\varphi/\sqrt{3}}+R_2 e^{2\varphi/\sqrt{3}}\right) ,
\ee
where
\begin{align}\label{params4}
    & V_0=\frac{\sqrt{2}C_2C_w}{\langle\vol\rangle^3e^{\langle\varphi\rangle/\sqrt{3}}},\qquad R_0=\frac{C_3e^{-\langle\varphi\rangle/\sqrt{3}}}{\sqrt{2}C_2C_w},\qquad \frac{R_1}{R_0}=\frac{\sqrt{2}e^{\sqrt{3}\langle\varphi\rangle}}{\langle\vol\rangle},\nonumber\\
    & \frac{R_2}{R_0}\sim\frac{C_2C_we^{4\langle\varphi\rangle/\sqrt{3}}}{C_3\langle\vol\rangle},\qquad C_{up}=1-R_0-R_1-R_2.
\end{align}
{}Following the usual lore, setting the vacuum expectation value of $\varphi$ to zero, we demand $R_0=\frac{1+R_1+2R_2}{2}\approx\frac{1}{2}$. $\delta$ has the usual meaning, as discussed in the case of the previous three models.

\section{WI setup}\label{sec:3}

Let us now present the WI implementation for the four FI types of potentials discussed in the previous section.

\subsection{Background dynamics of WI}

{}For WI, at the background level, one is interested in the cosmological evolution of both the scalar field, $\phi$, and the radiation energy density, $\rho_r$, of the thermal bath. The appropriate background evolution equations, in the presence of a dissipation coefficient $\Upsilon$, are: 
\begin{eqnarray}
&&\ddot \phi + (3 H + \Upsilon) \dot \phi + V_{,\phi}=0,
\label{eq.phi}
\\
&&\dot \rho_r + 4 H \rho_r = \Upsilon \dot \phi^2,
\label{eq.rho}
\end{eqnarray}
where  $V_{,\phi} = dV(\phi)/d\phi$, $H$ is the Hubble parameter:
\begin{equation}
H^2 = \frac{1}{3 M_{\rm Pl}^2} \left( \frac{\dot \phi^2}{2} + V + \rho_r \right).
\label{hubble}
\end{equation}
{}For a thermalized radiation bath, $\rho_r$ is related to the temperature as 
$\rho_r=C_R T^4$ with $C_R=\pi^2g_{\star}/30$ and $g_\star$ is the effective number of degrees of freedom for the radiation.
Here, we will be assuming $g_\star=106.75$, e.g. like for the standard model. 
Just like in  cold inflation (CI), accelerated inflationary regime ends when the Hubble slow-roll coefficient is $\epsilon_H=1$, where:
\begin{equation}
\epsilon_H = - \frac{\dot H}{H^2} \simeq \frac{\epsilon_V}{1+Q},
\label{epsH}
\end{equation}
with
\begin{equation}
\epsilon_V = \frac{ M_{\rm Pl}^2 }{2} \left( \frac{ V_{,\phi} }{V} \right)^2,
\label{epsV}
\end{equation}
and $Q= \Upsilon/(3H)$ is the dissipation ratio commonly considered in the WI literature.
The value of $Q$ quantifies how much inflaton energy has dissipated to radiation bath during inflation when compared
to the rate of expansion. $Q\gg 1$ denotes the strong dissipative regime of WI, 
while $Q\ll 1$ denotes the weak dissipative regime. 

\subsection{Perturbation spectrum in WI and observational quantities}

The primordial scalar of curvature power spectrum for a WI model can be generically expressed as~\cite{Ramos:2013nsa,Kamali:2023lzq}:
\be
P_{\mathcal{R}}\simeq\left(\frac{H^2}{2\pi \dot{\phi}}\right)^2\left(1+2n_{\star}+\frac{2\sqrt{3}Q}{\sqrt{3+4\pi Q}}
\frac{T}{H}\right)G(Q)\Bigr|_{k_*=aH},
\label{power_spec_gen}
\ee
where $n_{\star}$ quantifies the possibility of the thermalization of the inflaton perturbations, with $n_*=0$ 
denoting no-thermalization, and it else assumes the Bose-Einstein distribution, $n_{\star}\equiv n_{BE}=1/(e^{H/T}-1)$. 
$G(Q)$ in \eqref{power_spec_gen}, on the other hand, is dependent on the explicit form of the dissipation coefficient $\Upsilon$, 
such as its functional dependence on temperature and inflaton amplitude, etc. 
The function $G(Q)$ is generally understood to depend primarily on the form of the dissipation coefficient~\cite{Montefalcone:2023pvh}. 
However, as shown in the recent study~\cite{Rodrigues:2025neh}, when $n_{\star}=0$, $G(Q)$ could exhibit some dependence on the potential form for a range of $Q-$values. 
$G(Q)$ is determined by taking a ratio of the numerical power spectrum calculated by solving the coupled perturbation equations of both scalar field and the thermal fluctuations with the analytical one of \eqref{power_spec_gen} when written in the absence of the multiplicative factor $G(Q)$ (see refs.~\cite{Kamali:2023lzq,Rodrigues:2025neh}). 

\begin{center}
\begin{figure*}[!bth]
\subfigure[]{\includegraphics[width=7.cm]{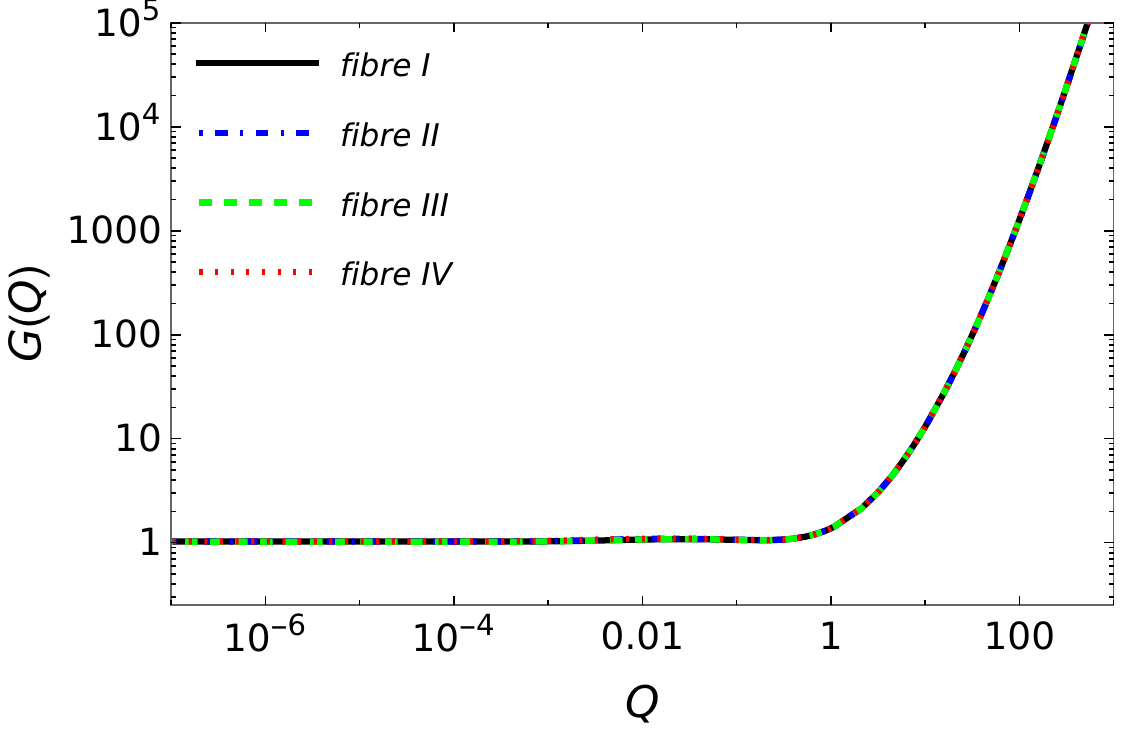}}
\subfigure[]{\includegraphics[width=7.cm]{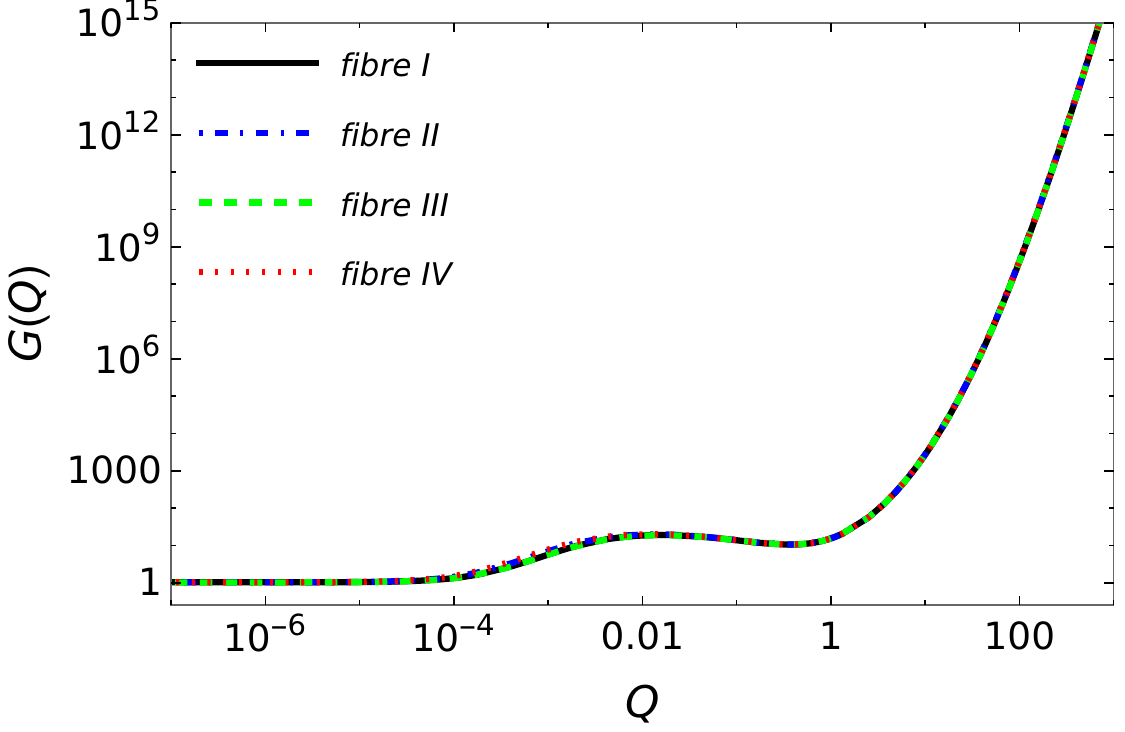}}
\caption{The $G(Q)$ function generated through  
\texttt{WI2easy}~\cite{Rodrigues:2025neh} for a dissipation coefficient eq.~(\ref{Ups}), with $f(T,\phi) =T^c \phi^p M_{\rm Pl}^{1-p-c}$ and considering the cases $c=1,p=0$ (panel a) and $c=3,p=0$ (panel b)
and for the parameters given in tables~\ref{tab1} and \ref{tab2} for the four type of fibre potentials. The nonthermal situation, $n_*=0$, in eq.~(\ref{power_spec_gen}) is considered in all the cases.}
\label{fig1}
\end{figure*}
\end{center}

The general expression for the dissipation coefficient can be written as
\begin{equation}
\Upsilon = C_{\Upsilon} f(T,\phi),
\label{Ups}
\end{equation}
where $C_{\Upsilon} $ is a dimensionless constant, typically depending on the microscopic parameters of the model
that give rise to the dissipation coefficient (such as coupling constants and other parameters of the model), while
$f(T,\phi)$ is some function of the temperature and inflaton amplitude. In most WI studies, $f(\phi,T)$ assumes a
simple form~\cite{BasteroGil:2010pb,BasteroGil:2012cm,Bastero-Gil:2016qru}
$f(T,\phi) =T^c \phi^p M_{\rm Pl}^{1-p-c}$. However, there are also WI model realizations in which the dependence on $T$ and $\phi$ can be
more intricate~\cite{Bastero-Gil:2019gao,Berghaus:2025dqi,ORamos:2025uqs}.
As already mentioned in the introduction, in the analysis we carry out here, we make use of two of the most well-known
forms of dissipation coefficient in WI, i.e., $\Upsilon \propto T$ and $\Upsilon \propto T^3$.
 {}For these two forms of dissipation coefficient, we can obtain the corresponding form for the function $G(Q)$ in eq.~(\ref{power_spec_gen})
and they are shown in fig.~\ref{fig1}. The results for $G(Q)$ obtained for the four types of fibre inflation potentials are almost indistinguishable, with $G(Q)$ mostly depending only on the form of the dissipation coefficient. In all of our numerical analysis, we use a spline interpolation of the numerical data for $G(Q)$ for each form for the dissipation coefficient. Then, we used these numerically interpolated functions for
$G(Q)$ in the power spectrum eq.~(\ref{power_spec_gen}).

Using the scalar power spectrum \eqref{power_spec_gen}, the spectral index $n_s$ can be calculated at the Hubble-exit point $(k_{\star})$ as:
\be
n_s-1=\frac{d\ln P_{\mathcal{R}}(k)}{d\ln \,k}\bigg|_{k\to k_{\star}}
\ee
where we can apply the following simplification, $d\ln k=(d\ln k/dN)d N$, $N=\ln a$. 
The running of the spectral index and tensor-to-scalar ratio are given, respectively, by
\begin{align}
\alpha_s&=\frac{d\, n_s(k)}{d\,\ln \,k}\bigg|_{k\to k_{\star}},\\
r&=\frac{P_{\mathcal{T}}}{P_{\mathcal{R}}},
\end{align}
where $P_{\mathcal{T}}=2H^2/(\pi^2M_{\rm Pl}^2)$ is the tensor spectrum.

Note that WI does not face the uncertainty in measuring the number of {\it e}-folds of inflation, $N_\star$, between the moment the 
relevant scales with wavenumber $k_\star$ leave the Hubble radius and reenter around today. 
Since there is no requirement for reheating, as the inflaton continues to dissipate to radiation throughout inflation, 
$N_\star$ can be directly obtained from\footnote{This is obtained as usual~\cite{Liddle:2003as}, by equating 
$k_\star=a_\star H_\star$ with $k_0=a_0H_0$,
with subindex "0" meaning quantities evaluated at present time, assuming $a_\star/a_{end}=e^{-(N_\star)}$ with $a_\star =1$, and applying entropy conservation results.}~\cite{Das:2020xmh}:
\be\label{N_starcal}
\dfrac{k_\star}{a_0 H_0}=e^{-N_{\star}}\left[\dfrac{43}{11g_s(T_{end})}\right]^{1/3}\dfrac{a_{end}}{a_{reh}}\dfrac{T_0}{T_{end}}\dfrac{H_{\star}}{H_0},
\ee
where $g_s(T_{end})$ is the entropy number of degrees of freedom at the end of inflation, we use $k_{\star}=0.05\,\mathrm{Mpc^{-1}}$, $a_0=1$ is the scale factor today, and $H_0=67.66\,\mathrm{km \,s^{-1}\,Mpc^{-1}}$ (the value obtained from Planck collaboration \cite{Planck:2018vyg}, TT,TE,EE-lowE+lensing+BAO $68\%$ CL) and $T_0=2.725\,\mathrm{K}\simeq 2.349\times 10^{-13}\,\mathrm{GeV}$ is the present day value of the cosmological microwave background (CMB) temperature. The parameter $a_{end}/a_{reh}$ gives the estimate of the number of {\it e}-folds that lasts between the end of inflation and the commencement of radiation domination. This is determined when the equation of state parameter changes its value from $-1$ during inflation to the value $1/3$, which is typical of radiation domination. In WI, this period typically lasts about $1-6$ efolds. In all of our numerical evaluations we use \texttt{WI2easy} \footnote{\texttt{WI2easy} is based on the methods developed in refs.~\cite{Ballesteros:2022hjk,Ballesteros:2023dno} for threating WI perturbations.}, a precision computational tool for the analysis of the background dynamics of WI and perturbation~\cite{Rodrigues:2025neh}. 

\section{Predictions from the warmer side}\label{sec:numerical_results}

In this section, we present our numerical analysis considering the four types of FI potential discussed in section~\ref{section_fourFibre}. 
Each potential model is also analyzed with the dissipation coefficients explained in the previous section, namely, 
$\Upsilon = C_\Upsilon T$ and $\Upsilon = C_\Upsilon T^3/M_{\rm Pl}^2$.
{}For each model, we analyze their predictions in both weak and strong dissipative regime. 
We find that FI can support both regimes, offering an attempt to  solve the longstanding issue in FI in the context of CI, where the field excursion consistently exceeds the Planck scale. We present the numerical results in tables~\ref{tab1}-\ref{tab3}. 

\subsection{Fibre inflation in weak dissipation}

In tables~\ref{tab1} and \ref{tab2}, row-wise we have the dimensionless dissipation constant $C_{\Upsilon}$, the potential normalization $V_0$, the 
potential free parameters $R$ from type-I and type-II and $(R_1,R_2)$ from type-III and type-IV, 
the value of the temperature, inflaton amplitude and inflaton time derivative all at Hubble-exit ($N_\star$), $T_\star$, $\phi_\star$
and $\dot \phi_\star$, respectively. Next, we present $N_{\star}$, 
which tells us the difference between Hubble exit and end of inflation. 
We present the scale of the potential $V^{1/4}$, value of the dissipation ratio $Q$, the ratio $T/H$, the 
potential slow-roll parameters $\epsilon_V$, eq.~(\ref{epsV}) and also the slow-roll coefficient
$\eta_V = M_{\rm Pl}^2 V_{,\phi\phi}/V$, the spectral index $n_s$, the tensor-to-scalar ratio $r$, the
running of the spectral index $\alpha_s$, all calculated at the Hubble-exit point, and the variation of the inflaton amplitude in the range from Hubble-exit up to the end of inflation. 

\begin{table}[H]
\centering
\begin{tabular}{|c|c|c| c|c|}

    \hline 

   &     \multicolumn{2}{|c|}{\cellcolor[gray]{0.9} \textbf{Type-I}  \cite{Cicoli:2008gp}} &     \multicolumn{2}{|c|}{\cellcolor[gray]{0.9} \textbf{Type-II} \cite{Cicoli:2016chb}} \\ \hline
     \cellcolor[gray]{0.9} Dissipation type   &  $\sim \,T$     &  $\sim \,T^3$   & $\sim \,T$  &  $\sim \,T^3$   \\ 
     \hline
         \multicolumn{5}{|c|}{\cellcolor[gray]{1.0} \textbf{Parameters}} \\ \hline
   \cellcolor[gray]{0.9}  $C_{\Upsilon}$  & $2.35\times 10^{-5}$ &  $2.05\times 10^5$ &  $2.35\times 10^{-5}$ &  $1.82 \times 10^5$ \\ 
     \cellcolor[gray]{0.9}  $V_0/M_{\rm Pl}^4$  & $5.82\times 10^{-11}$ &  $7.33\times 10^{-11}$ & $2\times 10^{-10}$ &  $2.52\times 10^{-10}$ \\ 
          \cellcolor[gray]{0.9}  $R$  & $3.05\times 10^{-8}$ &  $3.05\times 10^{-8}$ & $3.05\times 10^{-7}$ &  $3.05\times 10^{-7}$ \\ 
 
\hline \hline
        \multicolumn{5}{|c|}{\cellcolor[gray]{1.0} \textbf{Initial conditions (at Hubble-exit)}} \\ \hline

      \cellcolor[gray]{0.9}  $T_\star/M_{\rm Pl}$   & $9.48\times 10^{-6}$  &  $1.03\times 10^{-5}$ &  $9.96\times 10^{-6}$ &  $1.1\times 10^{-5}$\\

      \cellcolor[gray]{0.9}  $\phi_\star/M_{\rm Pl}$   & 5.92 &  5.72 & $6.50$  & $6.28$  \\

      \cellcolor[gray]{0.9}  $\dot\phi_\star/M_{\rm Pl}^2$   & $-1.96\times 10^{-7}$  &  $-2.47 \times 10^{-7}$ &  $-2.16\times 10^{-7}$ &  $-2.80\times 10^{-7}$ \\

            \hline    
        \hline
     \multicolumn{5}{|c|}{\cellcolor[gray]{1.0} \textbf{Horizon exit}} \\ \hline

      \cellcolor[gray]{0.9}  $N_\star$   & $57.94$  &  $56.10$ &  $58.04$ &  $56.18$ \\

           \hline  \hline
     \multicolumn{5}{|c|}{\cellcolor[gray]{1.0} \textbf{Values of the dynamical variables at} $N_\star$ } \\ \hline
      \cellcolor[gray]{0.9}  $V(\phi_\star)^{1/4}/M_{\rm Pl}$    &  $3.60\times 10^{-3}$ &  $3.80\times 10^{-3}$ &  $3.68\times 10^{-3}$ &  $3.93\times 10^{-3}$ \\ 
     \cellcolor[gray]{0.9}  $Q_{\star}$    & $9.97\times 10^{-6}$  &  $8.97\times 10^{-6}$ &  $9.97\times 10^{-6}$ &  $8.99\times 10^{-6}$ \\
      \cellcolor[gray]{0.9}  $T_{\star}/H_{\star}$ & $1.27$  &  $1.23$ &  $1.27$ &  $1.23$ \\
      
                  \hline \hline
     \multicolumn{5}{|c|}{\cellcolor[gray]{1.0} \textbf{Slow-roll parameters}} \\ \hline
      \cellcolor[gray]{0.9}  $\epsilon_{V_{\star}}$    & $3.45\times 10^{-4}$  &  $4.43\times 10^{-4}$ &  $3.83\times 10^{-4}$ &  $4.98\times 10^{-4}$ \\ 
     \cellcolor[gray]{0.9}  $\eta_{V_\star}$    & $-0.015$  &  $-0.017$ & $-0.015$ &  $-0.018$ \\

                        \hline \hline
     \multicolumn{5}{|c|}{\cellcolor[gray]{1.0} \textbf{Observable constraints from inflation}}
      \\ \hline
 
     \cellcolor[gray]{0.9}  $n_s$     & $0.968$ &  $0.965$ &  $0.967$ &  $0.962$ \\
      \cellcolor[gray]{0.9}  $r$   & $5.4\times 10^{-3}$  &  $6.7\times 10^{-3}$ &  $5.9\times 10^{-3}$ &  $7.6\times 10^{-3}$ \\

       \cellcolor[gray]{0.9}  $\alpha_s$ & $-5.4\times 10^{-4}$ &  $-4.2\times 10^{-4}$ &  $-4.9\times 10^{-4}$ &  $-7.6\times 10^{-4}$ \\  
            \cellcolor[gray]{0.9}  $\Delta \phi/M_{\rm Pl}$ & $5.33$  &  $5.49$ &  $5.81$ &  $6.02$ \\

      \hline 
 
\end{tabular}

\caption{The numerical values of the parameters and the associated cosmological quantities are obtained for the type-I-II fibre potentials in the weak dissipation regime, with $g_{\star} = 106.75$ across all models. In all realizations, the amplitude of the power spectrum is fixed at the pivot scale as $P_{\mathcal{R}}^{\star} = 2.1 \times 10^{-9}$. Inflaton is assumed to be non-thermalized with the radiation bath \ie $n_{\star}=0$.}
\label{tab1}

\end{table}

\begin{table}[H]
\centering
\begin{tabular}{|c|c|c| c|c|}

    \hline 

   &     \multicolumn{2}{|c|}{\cellcolor[gray]{0.9} \textbf{Type-III} \cite{Cicoli:2016xae}} &     \multicolumn{2}{|c|}{\cellcolor[gray]{0.9} \textbf{Type-IV} \cite{Bera:2024ihl}} \\ \hline
     \cellcolor[gray]{0.9} Dissipation type   &  $\sim \,T$     &  $\sim \,T^3$   & $\sim \,T$  &  $\sim \,T^3$   \\ 
     \hline
         \multicolumn{5}{|c|}{\cellcolor[gray]{1.0} \textbf{Parameters}} \\ \hline
   \cellcolor[gray]{0.9}  $C_{\Upsilon}$  & $2.35\times 10^{-5}$ &  $1.75 \times 10^5$ &  $2.35\times 10^{-5}$ &  $1.77 \times 10^5$ \\ 
     \cellcolor[gray]{0.9}  $V_0/M_{\rm Pl}^4$  & $7.25\times 10^{-11}$ &  $8.63\times 10^{-11}$ & $4.05\times 10^{-10}$ &  $5.18\times 10^{-10}$ \\ 
         \cellcolor[gray]{0.9}  $R_1$  & $ 10^{-6}$ & $ 10^{-6}$ & $2\times 10^{-7}$ &  $2\times 10^{-7}$ \\ 
 
    \cellcolor[gray]{0.9}  $R_2$  & $7\times 10^{-4}$ &  $7\times 10^{-4}$ & $ 2\times 10^{-7}$ &  $2\times 10^{-7}$ \\

            \hline    
        \hline

        \multicolumn{5}{|c|}{\cellcolor[gray]{1.0} \textbf{Initial conditions (at Hubble-exit)}} \\ \hline

      \cellcolor[gray]{0.9}  $T_\star/M_{\rm Pl}$   & $1.06\times 10^{-5}$  &  $1.13\times 10^{-5}$ &  $1.01\times 10^{-5}$ &  $1.11\times 10^{-5}$ \\

      \cellcolor[gray]{0.9}  $\phi_\star/M_{\rm Pl}$   & $6.02$ & $5.82$  &  $6.52$ & $6.20$  \\

      \cellcolor[gray]{0.9}  $\dot\phi_\star/M_{\rm Pl}^2$    & $-2.46\times 10^{-7}$  &  $-2.92\times 10^{-7}$ &  $-2.26\times 10^{-7}$ &  $-2.88\times 10^{-7}$ \\

 \hline \hline
      
     \multicolumn{5}{|c|}{\cellcolor[gray]{1.0} \textbf{Horizon exit}} \\ \hline

      \cellcolor[gray]{0.9}  $N_\star$   & $58.13$  &  $56.17$ &  $58.07$ &  $56.18$ \\

           \hline  \hline
     \multicolumn{5}{|c|}{\cellcolor[gray]{1.0} \textbf{Values of the dynamical variables at} $N_\star$ } \\ \hline
      \cellcolor[gray]{0.9}  $V(\phi_\star)^{1/4}/M_{\rm Pl}$    &  $3.80\times 10^{-3}$ &  $3.97\times 10^{-3}$ &  $3.72\times 10^{-3}$ &  $3.95\times 10^{-3}$ \\ 
     \cellcolor[gray]{0.9}  $Q_{\star}$    & $9.98\times 10^{-6}$  &  $9.22\times 10^{-6}$ &  $9.97\times 10^{-6}$ &  $9.04\times 10^{-6}$ \\
      \cellcolor[gray]{0.9}  $T_{\star}/H_{\star}$ & $1.27$  &  $1.23$ &  $1.27$ &  $1.23$ \\
      
                  \hline \hline
     \multicolumn{5}{|c|}{\cellcolor[gray]{1.0} \textbf{Slow-roll parameters}} \\ \hline
      \cellcolor[gray]{0.9}  $\epsilon_{V_\star}$    & $4.35\times 10^{-4}$  &  $5.17\times 10^{-4}$ &  $4.01\times 10^{-4}$ &  $5.11\times 10^{-4}$ \\ 
     \cellcolor[gray]{0.9}  $\eta_{V_\star}$    & $-0.011$  &  $-0.013$ & $-0.014$ &  $-0.017$ \\

                    \hline    \hline
     \multicolumn{5}{|c|}{\cellcolor[gray]{1.0} \textbf{Observable constraints from inflation}}
      \\ \hline
 
     \cellcolor[gray]{0.9}  $n_s$     & $0.975$ &  $0.971$ &  $0.969$ &  $0.964$ \\
      \cellcolor[gray]{0.9}  $r$   & $6.7\times 10^{-3}$  &  $8.0 \times 10^{-3}$ &  $6.2\times 10^{-3}$ &  $7.9\times 10^{-3}$ \\
            \cellcolor[gray]{0.9}  $\alpha_s$ & $-5.9\times 10^{-4}$ &  $-8.0\times 10^{-4}$ &  $-7.1\times 10^{-4}$ &  $-8.2\times 10^{-4}$ \\  
            
            \cellcolor[gray]{0.9}  $\Delta \phi/M_{\rm Pl}$ & $5.43$  &  $5.58$ &  $5.82$ &  $6.03$ \\ 
   
      \hline 
 
\end{tabular}

\caption{The numerical values of the parameters and the associated cosmological quantities are obtained for the type-III-IV fibre potentials in the weak dissipation regime, with $g_{\star} = 106.75$ across all models. In all realizations, the amplitude of the power spectrum is fixed at the pivot scale as $P_{\mathcal{R}}^{\star} = 2.1 \times 10^{-9}$. Inflaton is assumed to be non-thermalized with the radiation bath \ie $n_{\star}=0$.}
\label{tab2}

\end{table}

\begin{center}
\begin{figure}[!bth]
\subfigure[$\Upsilon=C_\Upsilon T$]{\includegraphics[width=7.5cm]{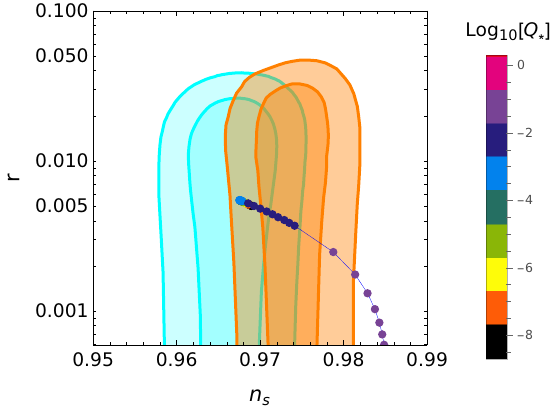}}
\subfigure[$\Upsilon=C_\Upsilon T^3/M_{\rm Pl}^2$]{\includegraphics[width=7.5cm]{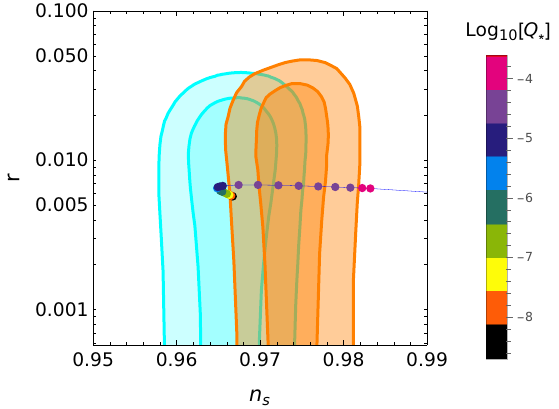}}
\caption{The one-sigma and two-sigma constrains from PlanckTT,TE,EE+lowE+lensing
+BK18+BAO (cyan shaded regions) and the combined P-ACT-LB results (orange shaded regions) for $r$ and $n_s$, together with the results for the warm FI type-I model with $R=3.05\times 10^{-8}$, for a dissipation coefficient with a linear and cubic dependencies on the temperature (panels a and b, respectively).}    
\label{fig2}
\end{figure}
\end{center}

The values of  $(n_s,r)$ lie well within the BICEP, Keck Array and Planck combined data~\cite{BICEP:2021xfz}, demonstrating that FI models of type-I-IV  with either a linear or a cubic dissipation function are viable candidates for WI. The value of $Q_{\star}$ confirms that we reside in the weak regime. One crucial thing to note here is that the model parameter $R$ for fibre type-I and type-II matches the value used in CI as can be found in 
refs.~\cite{Cicoli:2008gp,Cicoli:2016chb,Cicoli:2016xae,Bera:2024ihl}. 
By using  values of $Q$ larger than the ones shown in tables~\ref{tab1} and \ref{tab2}, $Q \gtrsim 10^{-5}$, we find that the spectral tilt $n_s$ 
starts to deviate
from  its central value~\cite{Planck:2018vyg}  $n_s=0.965 \pm 0.004$ and tends to become bluer (i.e., leading to larger values for $n_s$
as $Q$ increases), leaving the two-sigma constraint region).
Therefore, by choosing parameters $R, \; R_1, \; R_2$ for the FI potentials that are close to the ones used in CI, we find that WI is only compatible
with the observations in the weak dissipative regime of WI ($Q \lesssim 10^{-4}$).
Note, however, that the recent data from ACT~\cite{ACT:2025tim, ACT:2025fju} favors
slightly larger values of $n_s$ than earlier results from ref.~\cite{BICEP:2021xfz}. As a consequence, this allows us to push
$Q$ to slightly larger values than the ones allowed from the earlier Planck data. This trend was also explicitly shown in~\cite{Berera:2025vsu} for other
types of models in the WI realm. 
{}For illustration, we show in fig.~\ref{fig2} the results for the fibre type-I model in the plane ($n_s,r$) for the linear and cubic dependencies in the temperature. The contours  show the one-sigma and two-sigma constraints
from the combined datasets PlanckTT,TE,EE+lowE+lensing+BK18+BAO~\cite{BICEP:2021xfz} (cyan shaded regions) and  the recent data from ACT (P-ACT-LB-BK18 combined datasets)~\cite{ACT:2025tim}
(orange shaded regions).

Being in the weak regime, the value of $r$ is at the verge of detector sensitivity and $\Delta\phi$ is still super-Planckian, typical 
also of FI in the cold regime. {}Finally, observe that the value of $T/H\sim\mathcal{O}(1)$, i.e. it is at the boundary of being warm
(with the WI regime defined by $T/H > 1$). This is often the case for WI in the weak regime~\cite{Benetti:2016jhf}. 
Inflationary predictions remain qualitatively unchanged even in the presence of thermalization of the inflaton with the radiation bath, i.e. $n_{\star}\neq 0$.
This is why we do not present results for the thermal case for the inflaton perturbations ($n_*\equiv n_{BE}$). 
 
To be able to reach strong dissipation, we allow the potential parameters $R, \; R_1, \; R_2$  to be free parameters. In particular, we find that 
increasing $R, \; R_1, \; R_2$  with respect to their typical CI values can allow us to explore larger values of $Q$.
Next, assuming that $R, \; R_1, \; R_2$ are free parameters, we explore their compatibility with observations when increasing $Q$.
In particular, we will focus next on the cubic dissipation coefficient, $\Upsilon = C_\Upsilon T^3/M_{\rm Pl}^2$, which is well known to 
allow to produce consistent WI results for other types of primordial inflaton potentials in the strong regime of WI~\cite{Das:2020xmh,Das:2022ubr}.
In this case, the calculation of $P_{\mathcal{R}}$ is strongly dependent on the dissipation coefficient
and on the function $G(Q)$ in eq.~(\ref{power_spec_gen}) becomes important. 
The results for $G(Q)$ are generated using the public code \texttt{WI2easy}~\cite{Rodrigues:2025neh}. The code automates the computation of $G(Q)$ based on the user-defined model and formulation of the dissipation coefficient in WI. The generated data undergoes spline interpolation within \texttt{WI2easy} to ensure smooth integration into subsequent calculations of observational quantities, such as power spectra or spectral indices. Using the numerical output provided by \texttt{WI2easy}, we derive results like those illustrated in fig.~\ref{fig2}.

\subsection{Fibre inflation in strong dissipation}

In table~\ref{tab3}, we quote the results of our numerical analysis for the cubic dissipation function when moving to
the strong dissipative regime of WI. We display the same observables as discussed in the case of weak dissipation. 
{}For all four types of FI discussed in section~\ref{section_fourFibre}, we are able to fall in the observational window of the BICEP, Keck Array and Planck's combined data~\cite{BICEP:2021xfz}. In all the examples presented below, although we can achieve $T\gg H$, 
we could only go up to $Q\sim \mathcal{O}(10)$. Another important point to note is that the values of the model parameters, such as the value of $R$ in type I and type II and $(R_1,R_2)$ for type III and IV, are much higher than their values in CI~\cite{Cicoli:2008gp,Cicoli:2016chb,Cicoli:2016xae,Bera:2024ihl}. 
Therefore, it is important to check the predictions from the string theory side. We know that $\{R^{\text{type I}},R^{\text{type II}},R_1,R_2\}$ 
are related to stringy parameters through eqs.~\eqref{params1}, \eqref{params2}, \eqref{params3}, \eqref{params4}. 
Let us now show that warm FI in the strong dissipative regime can be consistently embedded into string type IIB flux 
compactification\footnote{Note that the values of the FI potentials normalization $V_0$ values quoted are all also automatically
obtained from \texttt{WI2easy} by properly normalizing the inflaton potentials according to the CMB amplitude of the scalar power spectrum 
at the pivot scale.}:

\bea
\label{stringy_typeI}
\textbf{Type-I \eqref{params1}}\qquad & & \hskip-0.5cm {V}_0 \sim \mathcal{O} (10^{-20}), \quad { R} \sim \mathcal{O} (10^{-2}), \\
& & \hskip-0.5cm g_s \sim 0.1, \quad W_0\sim 72, \quad C_w \sim 0.095, \nonumber\\
& & \hskip-0.5cm C_1^{KK} \sim 1 , \quad C_2^{KK}\sim 2, \quad \langle \vol \rangle \sim \mathcal{O}(10^6). \nonumber
\eea

\bea
\label{stringy_typeII}
\textbf{Type-II \eqref{params2}}\qquad & & \hskip-0.5cm {V}_0 \sim \mathcal{O} (10^{-16}), \quad { R} \sim \mathcal{O} (10^{-2}), \\
& & \hskip-0.5cm g_s \sim 0.001, \quad W_0\sim 72, \quad C_w \sim 0.095, \quad |\lambda|\sim 10^{-4},\nonumber\\
& & \hskip-0.5cm \alpha\sim 1.5,\quad \Pi_1 \sim \mathcal{O}(10) , \quad \Pi_2\sim \mathcal{O}(1) , \quad \langle \vol \rangle \sim \mathcal{O}(10^4). \nonumber
\eea

\bea
\label{stringy_typeIII}
\textbf{Type-III \eqref{params3}}\qquad & & \hskip-0.5cm {V}_0 \sim \mathcal{O} (10^{-21}), \quad { R_1} \sim \mathcal{O} (10^{-1}), \quad { R_2} \sim \mathcal{O} (10^{-3}), \\
& & \hskip-0.5cm g_s \sim 0.35, \quad W_0\sim 1.2, \quad C_w \sim 1, \quad |\lambda|\sim 10^{-2},\nonumber\\
& & \hskip-0.5cm \alpha\sim 1.5,\quad C_1^{KK} \sim 4 , \quad C_2^{KK}\sim 3 , \quad \langle \vol \rangle \sim \mathcal{O}(10^5). \nonumber
\eea

\bea
\label{stringy_typeIV}
\textbf{Type-IV \eqref{params4}}\qquad & & \hskip-0.5cm {V}_0 \sim \mathcal{O} (10^{-18}), \quad { R_1} \sim \mathcal{O} (10^{-1}), \quad { R_2} \sim \mathcal{O} (10^{-3}), \\
& & \hskip-0.5cm C_2C_w \sim \mathcal{O}(10^{-6}), \quad C_3\sim \mathcal{O}(10^{-4}),\nonumber\\
& & \hskip-0.5cm \langle \vol \rangle \sim \mathcal{O}(10^4),\quad \langle \varphi \rangle \sim 4.5. \nonumber
\eea

It is evident from the above parameter sets that fibre WI in $Q\gg 1$ can reside in the weak coupling and the large volume limit, 
satisfying the low-energy 4D supergravity approximation. Next, let us define various relevant mass scales:
\bea
M_s=\frac{g_s^{1/4}\sqrt{\pi}}{\sqrt{\mathcal{V}}}M_{\rm Pl},\quad M_{KK}=\frac{\sqrt{\pi}}{\mathcal{V}^{2/3}}M_{\rm Pl},\quad m_{3/2}=\sqrt{\kappa}\frac{|W_0|}{\mathcal{V}}M_{\rm Pl},\quad \kappa=\frac{g_s}{8\pi}.
\eea
In order to trust the underlying effective low energy supergravity approximation, we need to maintain the following hierarchy of mass scales,
\begin{equation}\label{hierarchy1}
    m_{3/2}<M_{KK}<M_s<M_{\rm Pl}.
\end{equation}
{}For values of $W_0\sim \mathcal{O}(10),\,\langle\vol\rangle \sim \mathcal{O}(10^5),\,g_s \sim \mathcal{O}(10^{-2})$, one can easily satisfy the hierarchy of \eqref{hierarchy1}.

\begin{table}[H]
\centering
\begin{tabular}{|c|c|c| c|c|}

    \hline 

   &    {\cellcolor[gray]{0.9} \textbf{Type-I} \cite{Cicoli:2008gp} } &     {\cellcolor[gray]{0.9} \textbf{Type-II} \cite{Cicoli:2016chb}} &     {\cellcolor[gray]{0.9} \textbf{Type-III} \cite{Cicoli:2016xae}} &     {\cellcolor[gray]{0.9} \textbf{Type-IV} \cite{Bera:2024ihl}} \\ \hline
         \multicolumn{5}{|c|}{\cellcolor[gray]{1.0} \textbf{Parameters}} \\ \hline
   \cellcolor[gray]{0.9}  $C_{\Upsilon}$  & $1.02\times 10^{9}$ &  $6.03\times 10^7$ &  $1.94\times 10^{9}$ &  $2.50\times 10^8$ \\ 
     \cellcolor[gray]{0.9}  $V_0/M_{\rm Pl}^4$  & $6.83\times 10^{-20}$ &  $2.43\times 10^{-16}$ & $9.46\times 10^{-21}$ &  $5.86\times 10^{-18}$ \\ 
     \cellcolor[gray]{0.9}  $R$  & $0.06$ &  $0.06$ & --- &  --- \\ 

         \cellcolor[gray]{0.9}  $R_1$  & --- & --- & $0.1$ &  $0.1$ \\ 
 
    \cellcolor[gray]{0.9}  $R_2$  & --- &  --- & $0.001$ &  $0.001$ \\ 
      \cellcolor[gray]{0.9}  $\delta$  & $5.79\times 10^{-4}$ &  $8.5\times 10^{-4}$ & $0.0016$ &  $0.00437$ \\

            \hline             \hline

     \multicolumn{5}{|c|}{\cellcolor[gray]{1.0} \textbf{Initial conditions (at Hubble-exit)}} \\ \hline

        \cellcolor[gray]{0.9}  $\phi_\star/M_{\rm Pl}$   &  $2.38$ & $3.91$  &  $2.20$ &  $3.22$ \\

      \cellcolor[gray]{0.9}  $\dot\phi_\star/M_{\rm Pl}^2$    & $-8.93\times 10^{-12}$  &  $-5.74\times 10^{-10}$ &  $-3.21\times 10^{-12}$ &  $-6.45\times 10^{-11}$ \\

      \cellcolor[gray]{0.9}  $T_\star/M_{\rm Pl}$   & $2.27\times 10^{-6}$  &  $1.37\times 10^{-5}$ &  $1.45\times 10^{-6}$ &  $5.46\times 10^{-6}$ \\

           \hline            \hline

               \multicolumn{5}{|c|}{\cellcolor[gray]{1.0} \textbf{Horizon exit}} \\ \hline

      \cellcolor[gray]{0.9}  $N_\star$   & $51.45$  &  $53.30$ &  $51.04$ &  $52.40$ \\
                  \hline   
                             \hline

     \multicolumn{5}{|c|}{\cellcolor[gray]{1.0} \textbf{Values of the dynamical variables at} $N_\star$ } \\ \hline
      \cellcolor[gray]{0.9}  $V(\phi_\star)^{1/4}/M_{\rm Pl}$    &  $2.1\times 10^{-5}$ &  $1.34\times 10^{-4}$ &  $1.30\times 10^{-5}$ &  $4.85\times 10^{-5}$ \\ 
     \cellcolor[gray]{0.9}  $Q_{\star}$    & $15.5$  &  $5.0$ &  $20.0$ &  $10.0$ \\
      \cellcolor[gray]{0.9}  $T_{\star}/H_{\star}$ & $8856.6$  &  $1324.2$ &  $14723.0$ &  $4017.4$ \\
      
                  \hline \hline
     \multicolumn{5}{|c|}{\cellcolor[gray]{1.0} \textbf{Slow-roll parameters}} \\ \hline
      \cellcolor[gray]{0.9}  $\epsilon_{V_\star}$    & $0.167$  &  $0.056$ &  $0.24$ &  $0.14$ \\ 
     \cellcolor[gray]{0.9}  $\eta_{V_\star}$    & $0.324$  &  $0.103$ & $0.44$ &  $0.25$ \\

                        \hline \hline
     \multicolumn{5}{|c|}{\cellcolor[gray]{1.0} \textbf{Observable constraints from inflation}}
      \\ \hline
 
     \cellcolor[gray]{0.9}  $n_s$     & $0.966$ &  $0.975$ &  $0.967$ &  $0.968$ \\
      \cellcolor[gray]{0.9}  $r$   & $6.31\times 10^{-12}$  &  $7.19\times 10^{-9}$ &  $9.34\times 10^{-13}$ &  $1.78\times 10^{-10}$ \\

            \cellcolor[gray]{0.9}  $\alpha_s$ & $0.0035$ &  $0.0021$ &  $0.0027$ &  $0.0016$ \\  

            \cellcolor[gray]{0.9}  $\Delta \phi/M_{\rm Pl}$ & $2.33$  &  $3.85$ &  $2.16$ &  $3.26$ \\

      \hline 
 
\end{tabular}

\caption{The numerical values of the parameters and the associated cosmological quantities are obtained for the type-I-IV fibre potentials in the strong dissipation regime with the dissipation coefficient 
$\Upsilon = C_\Upsilon T^3/M_{\rm Pl}^2$. In all realizations, the amplitude of the power spectrum is fixed at the pivot scale as $P_{\mathcal{R}}^{\star} = 2.1 \times 10^{-9}$. The inflaton is assumed to be non-thermalized with the radiation bath \ie $n_{\star}=0$ and we have also considered $g_{\star} = 106.75$ across all models.}
\label{tab3}

\end{table}

\begin{center}
\begin{figure}[!bth]
\centerline{\includegraphics[width=7.5cm]{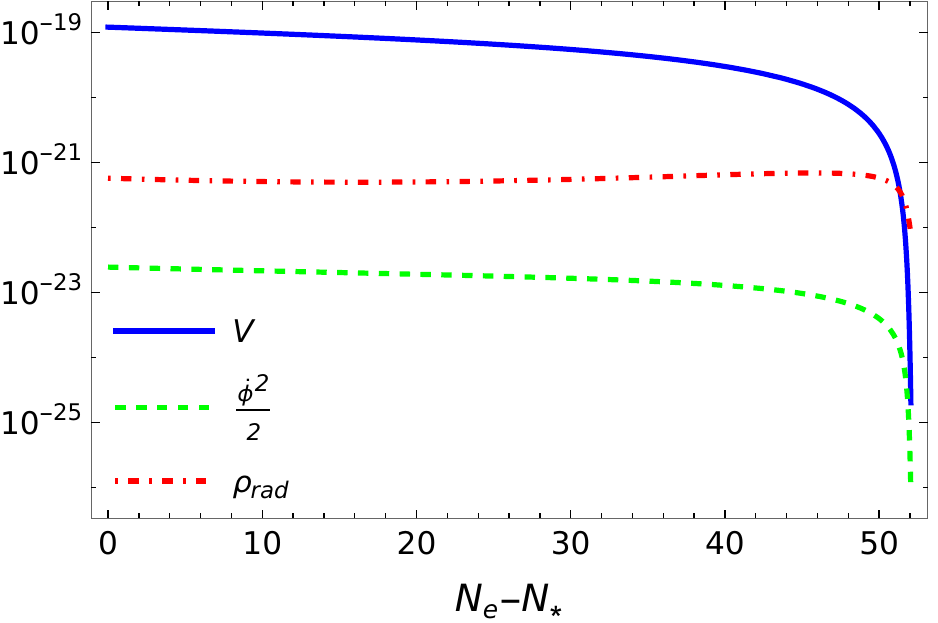}}
\caption{The evolution of the energy components in WI (in units of reduced Planck mass $M_{\rm Pl}$). We have considered the parameters given in table~\ref{tab3}, for the FI type-I model.}    
\label{fig3}
\end{figure}
\end{center}

It is useful to quote here the ratio $\rho_{r}/\rho_\phi$ between the radiation and inflaton’s energy densities during inflation. This can be easily obtained using the slow-roll approximations for the eqs.~(\ref{eq.phi}), (\ref{eq.rho}) and (\ref{hubble}),
\begin{eqnarray}
&& \dot \phi \simeq  - \frac{V_{,\phi}}{3H (1+Q)},
\label{eq.phi2}
\\
&& \rho_r \simeq 3Q \dot \phi^2/4,
\label{eq.rho2}
\end{eqnarray}
and
\begin{equation}
H^2 \simeq \frac{V}{3 M_{\rm Pl}^2},
\label{hubble2}
\end{equation}
from which we obtain
\begin{eqnarray}
\frac{\rho_r}{\rho_\phi} &=& \frac{\frac{\dot \phi^2}{2} + V}{\rho_r} \simeq \frac{ \frac{Q}{1+Q} \frac{\epsilon_H }{2}}{ 1+ \frac{\epsilon_H}{3(1+Q)}},
\label{rhoratio}
\end{eqnarray}
where we have used eq.~(\ref{epsH}).
Note that during WI we have an hierarchy $\dot \phi^2/2 \ll \rho_r \ll V$. This is well illustrated in the result for each of these energy components shown in fig.~\ref{fig3}, where we have used as an example the parameters for the fibre I potential given in table~\ref{tab3}.

{}From eq.~(\ref{rhoratio}), in the strong dissipation regimes $Q \gg 1$ and at the end of inflation, $\epsilon_H=1$, we have $\rho_r/\rho_\phi \sim 1/2$. Typically, radiation becomes dominant over
the inflaton energy density around one to three {\it e}-folds after the end of WI~\cite{Rodrigues:2025neh}.

One detail that we also observe from the results shown in table~\ref{tab3}, is that for strong dissipation, we need to assume large values of the model parameters. Therefore, it is necessary to introduce a new variable $\delta$ that takes care of the increase of the fiber potential to a Minkowski one. Note that, for type-II potential, the scalar field spectra is slightly less red-tilted, 
showing a larger value for $n_s$
compared to the other models. However, this is still consistent with the recent data from ACT~\cite{ACT:2025tim, ACT:2025fju}, which favor slightly larger values for the spectral tilt, $ n_s = 0.974 \pm 0.003$ 
(see also fig.~\ref{fig2}).
We also note from the results of table~\ref{tab3} that the models studied here in the strong dissipative regime consistently 
indicates a positive value for the running of the spectral tilt, $\alpha_s >0$, while the models in the weak dissipative regime,
tables~\ref{tab1} and \ref{tab2}, prefer a negative value for the running.
Although there is still a large uncertainty in the observational data with respect to the possible value for $\alpha_s$, the recent ACT analysis
seems to favor a slight positive value~\cite{ACT:2025tim}, $\alpha_s \equiv dn_s/d\ln k= 0.0062 \pm 0.0052$, while the latest Planck data favor a slightly negative (but still consistent with zero) result~\cite{Planck:2018vyg,Planck:2018jri}, $\alpha_s =  -0.0041\pm 0.0067$
(from the combined Planck CMB, lensing, and BAO data).

\begin{center}
\begin{figure}[!bth]
\centerline{\includegraphics[width=7.5cm]{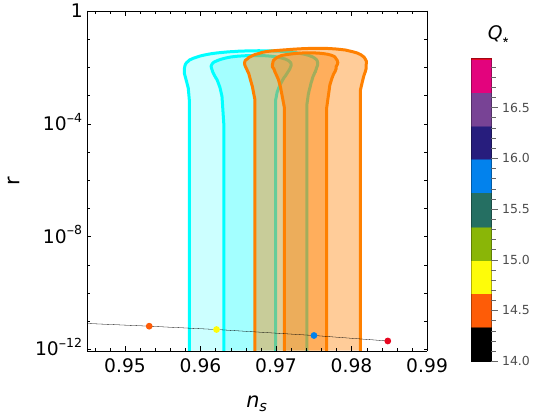}}
\caption{Same as in fig.~\ref{fig2}, but for the FI type-I model
for the dissipation coefficient with a cubic dependence on the temperature and for the strong dissipation regime. The parameters for the potential are given in table~\ref{tab3}.}    
\label{fig4}
\end{figure}
\end{center}

It is a known fact that WI in a strong dissipative regime~\cite{Motaharfar:2018zyb} is known to alleviate tensions from swampland programs~\cite{Vafa:2005ui,Obied:2018sgi, Agrawal:2018own, Garg:2018reu,Palti:2019pca}. With the help of this large dissipation ratio, it can give room to single field inflation models that are ruled out by the de Sitter-swampland conjectures~\cite{Motaharfar:2018zyb,Das:2018rpg,Das:2020xmh}. In addition, as can also be seen in table~\ref{tab3} that the value of $\Delta \phi$ tends to show a smaller value compared to FI in the weak dissipative regime (see tables~\ref{tab1} and \ref{tab2}). It shows a clear indication that increasing $Q$ further can push FI toward achieving a sub-Planckian field excursion. Although large $Q$ presents us with sub-Planckian field excursion --- making the model consistent with the distance swampland conjecture, increasing $Q$ and still residing in the observational window of Planck data comes at the cost of increasing the values of $(R,R_1,R_2)$, which may cause tension with the supergravity approximations and may push our WI realization away from the regime of perturbative control of type-IIB string theory. We point out that accommodating larger $Q$ will inevitably suppress the value
for the tensor-to-scalar ratio, $r$, even more, receding away from the detection of primordial gravitational waves (recalling that the most current up-to-date likelihood analysis of the CMB data~\cite{BICEP:2021xfz} leads to the constraint on
the tensor-to-scalar ratio to be $r < 0.036$ at 95\% confidence, with the next generation of CMB experiments possibly pushing this upper bound
to approximately $r \lesssim 10^{-4}$). 
To illustrate the results for the case of strong dissipation, we consider again the case shown in table~\ref{tab3} for the FI type-I model. Analogously to the results shown in fig. ~\ref{fig2}, we show the results for this case in the plane ($n_s,r$) in fig.~\ref{fig4}.
Similar results are also obtained for the other three types of FI.
Typically, we find that the larger the potential parameters $R, R_1, R_2$, the smaller is the range for $Q$ that falls within the observational window for $n_s$.

\section{Conclusion}\label{sec:conclusion}

In this paper, we have taken the first step towards embedding the well-known fibre inflationary model in type-IIB string theory in the framework of warm inflation. Since the original fibre inflationary scenario was proposed in ref.~\cite{Cicoli:2008gp}, three more variations of fibre inflation models have been introduced (see section~\ref{section_fourFibre} for details). Chronologically, we have named them as type-I-IV FI. We have demonstrated that fibre WI can withstand both strong and weak dissipation satisfying the observational data from both Planck and ACT. We show several benchmark examples in section~\ref{sec:numerical_results} to establish our claim as well as their clear connection to the string theory parameters. 

We have shown that by using typical CI FI potential parameters, we can only achieve WI in the weakly dissipative regime ($Q \lesssim 10^{-4}$).
Attempting to reach larger values of dissipation with these fixed string model parameters results in larger values for the spectral tilt, taking us
away from the observational window for $n_s$. 
To attempt to go to the strong dissipative regime of WI, we need to lose the constraints on the parameters of the FI potential we have studied
here. By assuming larger values for the potential parameters ${R,R_1,R_2}$ than the ones considered in CI, we show that we can achieve WI in the strong
regime. WI in the strong dissipative regime allows for instance
for the inflaton mass to be larger than the Hubble scale, $m_{\phi} > H$, and
the inflaton field excursion to remain below the Planck scale $\Delta\phi < M_{\rm Pl}$ \cite{Berera:1999ws,Berera:2003yyp}.
Having $Q\gg 1$ is very appealing from an effective field theory point of view since then the inflation model would be
protected from quantum gravity corrections arising in higher dimensional operators and free of uncontrollable infrared problems.  
Our results confirm that residing in the strong dissipative regime can in principle diminish the long-standing problem of super-Planckian inflaton field excursion 
in the realm of FI. Although this can be achieved by loosing the
constraints on the FI potential and, potentially taking us away from the perturbative control of string type-IIB flux compactification, 
because of the need of larger values for the model parameters, ${R,R_1,R_2}\gtrsim 1$.

{}Future studies could extend our analysis of FI potentials to alternative dissipation coefficients, 
such as the one derived and explored in ref.~\cite{Bastero-Gil:2019gao}. Such models may achieve strong dissipation
and sub-Planckian field excursions, while preserving two key features:
a red-tilted scalar spectrum consistent with CMB constraints and possibly also allowing
{}FI potential parameters compatible with string theory perturbativity bounds.
Notably, the concave profile ($V_{,\phi\phi}<0$) of FI potentials near their inflection points--where inflation occurs--enhances curvature perturbations when coupled to high-power in the temperature dissipative coefficients, like the one we have studied
here ($\Upsilon \propto T^3$). As shown in ref.~\cite{Ito:2025lcg}, this synergy can amplify secondary 
gravitational wave (GW) signals that might be detectable by next-generation interferometers (e.g., Einstein Telescope, DECIGO, etc). Similar GW signatures may arise in our studied models, contingent on the interplay between dissipation strength ($Q$) and the free parameters of the potential.
We hope to quantify these effects in a forthcoming work.

\section*{Acknowledgments}

D.C. thanks George K. Leontaris for useful discussions and comments.
R.O.R. acknowledges financial support by research grants from Conselho
Nacional de Desenvolvimento Cient\'{\i}fico e Tecnol\'ogico (CNPq),
Grant No. 307286/2021-5, and from Funda\c{c}\~ao Carlos Chagas Filho
de Amparo \`a Pesquisa do Estado do Rio de Janeiro (FAPERJ), Grant
No. E-26/201.150/2021.


\providecommand{\href}[2]{#2}\begingroup\raggedright\endgroup


\begin{thebibliography}{10}

\bibitem{Guth:1980zm}
A.~H.~Guth,
The Inflationary Universe: A Possible Solution to the Horizon and Flatness Problems,
Phys. Rev. D \textbf{23} (1981), 347-356
doi:10.1103/PhysRevD.23.347

\bibitem{Sato:1981ds}
K.~Sato,
Cosmological Baryon Number Domain Structure and the First Order Phase Transition of a Vacuum,
Phys. Lett. B \textbf{99} (1981), 66-70
doi:10.1016/0370-2693(81)90805-4

\bibitem{Albrecht:1982wi}
A.~Albrecht and P.~J.~Steinhardt,
Cosmology for Grand Unified Theories with Radiatively Induced Symmetry Breaking,
Phys. Rev. Lett. \textbf{48} (1982), 1220-1223
doi:10.1103/PhysRevLett.48.1220

\bibitem{Linde:1981mu}
A.~D.~Linde,
A New Inflationary Universe Scenario: A Possible Solution of the Horizon, Flatness, Homogeneity, Isotropy and Primordial Monopole Problems,
Phys. Lett. B \textbf{108} (1982), 389-393
doi:10.1016/0370-2693(82)91219-9

\bibitem{Linde:1983gd}
A.~D.~Linde,
Chaotic Inflation,
Phys. Lett. B \textbf{129} (1983), 177-181
doi:10.1016/0370-2693(83)90837-7

\bibitem{Baumann:2014nda}
D.~Baumann and L.~McAllister,
Inflation and String Theory,
Cambridge University Press, 2015,
ISBN 978-1-107-08969-3, 978-1-316-23718-2
doi:10.1017/CBO9781316105733
[arXiv:1404.2601 [hep-th]].

\bibitem{Candelas:1985en}
P.~Candelas, G.~T.~Horowitz, A.~Strominger and E.~Witten,
Vacuum configurations for superstrings,
Nucl. Phys. B \textbf{258} (1985), 46-74
doi:10.1016/0550-3213(85)90602-9

\bibitem{Greene:1993vm}
B.~R.~Greene, D.~R.~Morrison and M.~R.~Plesser,
Mirror manifolds in higher dimension,
Commun. Math. Phys. \textbf{173} (1995), 559-598
doi:10.1007/BF02101657
[arXiv:hep-th/9402119 [hep-th]].

\bibitem{Grana:2005jc}
M.~Grana,
Flux compactifications in string theory: A Comprehensive review,
Phys. Rept. \textbf{423} (2006), 91-158
doi:10.1016/j.physrep.2005.10.008
[arXiv:hep-th/0509003 [hep-th]].

\bibitem{Douglas:2006es}
M.~R.~Douglas and S.~Kachru,
Flux compactification,
Rev. Mod. Phys. \textbf{79} (2007), 733-796
doi:10.1103/RevModPhys.79.733
[arXiv:hep-th/0610102 [hep-th]].

\bibitem{Polchinski:1995mt}
J.~Polchinski,
Dirichlet Branes and Ramond-Ramond charges,
Phys. Rev. Lett. \textbf{75} (1995), 4724-4727
doi:10.1103/PhysRevLett.75.4724
[arXiv:hep-th/9510017 [hep-th]].

\bibitem{Marchesano:2007de}
F.~Marchesano,
Progress in D-brane model building,
Fortsch. Phys. \textbf{55} (2007), 491-518
doi:10.1002/prop.200610381
[arXiv:hep-th/0702094 [hep-th]].

\bibitem{Conlon:2005jm}
J.~P.~Conlon and F.~Quevedo,
Kahler moduli inflation,
JHEP \textbf{01} (2006), 146
doi:10.1088/1126-6708/2006/01/146
[arXiv:hep-th/0509012 [hep-th]].

\bibitem{Giddings:2001yu}
S.~B.~Giddings, S.~Kachru and J.~Polchinski,
Hierarchies from fluxes in string compactifications,
Phys. Rev. D \textbf{66} (2002), 106006
doi:10.1103/PhysRevD.66.106006
[arXiv:hep-th/0105097 [hep-th]].

\bibitem{Kachru:2003aw}
S.~Kachru, R.~Kallosh, A.~D.~Linde and S.~P.~Trivedi,
De Sitter vacua in string theory,
Phys. Rev. D \textbf{68} (2003), 046005
doi:10.1103/PhysRevD.68.046005
[arXiv:hep-th/0301240 [hep-th]].

\bibitem{Balasubramanian:2005zx}
V.~Balasubramanian, P.~Berglund, J.~P.~Conlon and F.~Quevedo,
Systematics of moduli stabilisation in Calabi-Yau flux compactifications,
JHEP \textbf{03} (2005), 007
doi:10.1088/1126-6708/2005/03/007
[arXiv:hep-th/0502058 [hep-th]].

\bibitem{Gukov:1999ya}
S.~Gukov, C.~Vafa and E.~Witten,
CFT's from Calabi-Yau four folds,
Nucl. Phys. B \textbf{584} (2000), 69-108
[erratum: Nucl. Phys. B \textbf{608} (2001), 477-478]
doi:10.1016/S0550-3213(00)00373-4
[arXiv:hep-th/9906070 [hep-th]].

\bibitem{Witten:1996bn}
E.~Witten,
Nonperturbative superpotentials in string theory,
Nucl. Phys. B \textbf{474} (1996), 343-360
doi:10.1016/0550-3213(96)00283-0
[arXiv:hep-th/9604030 [hep-th]].

\bibitem{vonGersdorff:2005bf}
G.~von Gersdorff and A.~Hebecker,
Kahler corrections for the volume modulus of flux compactifications,
Phys. Lett. B \textbf{624} (2005), 270-274
doi:10.1016/j.physletb.2005.08.024
[arXiv:hep-th/0507131 [hep-th]].

\bibitem{Berg:2004ek}
M.~Berg, M.~Haack and B.~Kors,
Loop corrections to volume moduli and inflation in string theory,
Phys. Rev. D \textbf{71} (2005), 026005
doi:10.1103/PhysRevD.71.026005
[arXiv:hep-th/0404087 [hep-th]].

\bibitem{Berg:2005ja}
M.~Berg, M.~Haack and B.~Kors,
String loop corrections to Kahler potentials in orientifolds,
JHEP \textbf{11} (2005), 030
doi:10.1088/1126-6708/2005/11/030
[arXiv:hep-th/0508043 [hep-th]].

\bibitem{Berg:2007wt}
M.~Berg, M.~Haack and E.~Pajer,
Jumping Through Loops: On Soft Terms from Large Volume Compactifications,
JHEP \textbf{09} (2007), 031
doi:10.1088/1126-6708/2007/09/031
[arXiv:0704.0737 [hep-th]].

\bibitem{Cicoli:2007xp}
M.~Cicoli, J.~P.~Conlon and F.~Quevedo,
Systematics of String Loop Corrections in Type IIB Calabi-Yau Flux Compactifications,
JHEP \textbf{01} (2008), 052
doi:10.1088/1126-6708/2008/01/052
[arXiv:0708.1873 [hep-th]].

\bibitem{Cicoli:2008va}
M.~Cicoli, J.~P.~Conlon and F.~Quevedo,
General Analysis of LARGE Volume Scenarios with String Loop Moduli Stabilisation,
JHEP \textbf{10} (2008), 105
doi:10.1088/1126-6708/2008/10/105
[arXiv:0805.1029 [hep-th]].

\bibitem{Antoniadis:2018hqy}
I.~Antoniadis, Y.~Chen and G.~K.~Leontaris,
Perturbative moduli stabilisation in type IIB/F-theory framework,
Eur. Phys. J. C \textbf{78} (2018) no.9, 766
doi:10.1140/epjc/s10052-018-6248-4
[arXiv:1803.08941 [hep-th]].

\bibitem{Antoniadis:2019rkh}
I.~Antoniadis, Y.~Chen and G.~K.~Leontaris,
Logarithmic loop corrections, moduli stabilisation and de Sitter vacua in string theory,
JHEP \textbf{01} (2020), 149
doi:10.1007/JHEP01(2020)149
[arXiv:1909.10525 [hep-th]].

\bibitem{Gao:2022uop}
X.~Gao, A.~Hebecker, S.~Schreyer and V.~Venken,
Loops, local corrections and warping in the LVS and other type IIB models,
JHEP \textbf{09} (2022), 091
doi:10.1007/JHEP09(2022)091
[arXiv:2204.06009 [hep-th]].

\bibitem{Becker:2002nn}
K.~Becker, M.~Becker, M.~Haack and J.~Louis,
Supersymmetry breaking and alpha-prime corrections to flux induced potentials,
JHEP \textbf{06} (2002), 060
doi:10.1088/1126-6708/2002/06/060
[arXiv:hep-th/0204254 [hep-th]].

\bibitem{Ciupke:2015msa}
D.~Ciupke, J.~Louis and A.~Westphal,
Higher-Derivative Supergravity and Moduli Stabilization,
JHEP \textbf{10} (2015), 094
doi:10.1007/JHEP10(2015)094
[arXiv:1505.03092 [hep-th]].

\bibitem{Burgess:2003ic}
C.~P.~Burgess, R.~Kallosh and F.~Quevedo,
De Sitter string vacua from supersymmetric D terms,
JHEP \textbf{10} (2003), 056
doi:10.1088/1126-6708/2003/10/056
[arXiv:hep-th/0309187 [hep-th]].

\bibitem{Cicoli:2015ylx}
M.~Cicoli, F.~Quevedo and R.~Valandro,
De Sitter from T-branes,
JHEP \textbf{03} (2016), 141
doi:10.1007/JHEP03(2016)141
[arXiv:1512.04558 [hep-th]].

\bibitem{Kachru:2002sk}
S.~Kachru, M.~B.~Schulz, P.~K.~Tripathy and S.~P.~Trivedi,
New supersymmetric string compactifications,
JHEP \textbf{03} (2003), 061
doi:10.1088/1126-6708/2003/03/061
[arXiv:hep-th/0211182 [hep-th]].

\bibitem{Crino:2020qwk}
C.~Crin{\`o}, F.~Quevedo and R.~Valandro,
On de Sitter String Vacua from Anti-D3-Branes in the Large Volume Scenario,
JHEP \textbf{03} (2021), 258
doi:10.1007/JHEP03(2021)258
[arXiv:2010.15903 [hep-th]].

\bibitem{Vafa:2005ui}
C.~Vafa,
The String landscape and the swampland,
[arXiv:hep-th/0509212 [hep-th]].

\bibitem{Cicoli:2024bxw}
M.~Cicoli, A.~Grassi, O.~Lacombe and F.~G.~Pedro,
Chiral global embedding of Fibre Inflation with $ \overline{\textrm{D}3} $ uplift,
JHEP \textbf{06} (2025), 090
doi:10.1007/JHEP06(2025)090
[arXiv:2412.08723 [hep-th]].

\bibitem{Cicoli:2023opf}
M.~Cicoli, J.~P.~Conlon, A.~Maharana, S.~Parameswaran, F.~Quevedo and I.~Zavala,
String cosmology: From the early universe to today,
Phys. Rept. \textbf{1059} (2024), 1-155
doi:10.1016/j.physrep.2024.01.002
[arXiv:2303.04819 [hep-th]].

\bibitem{Cicoli:2008gp}
M.~Cicoli, C.~P.~Burgess and F.~Quevedo,
Fibre Inflation: Observable Gravity Waves from IIB String Compactifications,
JCAP \textbf{03} (2009), 013
doi:10.1088/1475-7516/2009/03/013
[arXiv:0808.0691 [hep-th]].

\bibitem{Burgess:2014tja}
C.~P.~Burgess, M.~Cicoli, F.~Quevedo and M.~Williams,
Inflating with Large Effective Fields,
JCAP \textbf{11} (2014), 045
doi:10.1088/1475-7516/2014/11/045
[arXiv:1404.6236 [hep-th]].

\bibitem{Burgess:2016owb}
C.~P.~Burgess, M.~Cicoli, S.~de Alwis and F.~Quevedo,
Robust Inflation from Fibrous Strings,
JCAP \textbf{05} (2016), 032
doi:10.1088/1475-7516/2016/05/032
[arXiv:1603.06789 [hep-th]].

\bibitem{Cicoli:2016chb}
M.~Cicoli, D.~Ciupke, S.~de Alwis and F.~Muia,
$\alpha'$ Inflation: moduli stabilisation and observable tensors from higher derivatives,
JHEP \textbf{09} (2016), 026
doi:10.1007/JHEP09(2016)026
[arXiv:1607.01395 [hep-th]].

\bibitem{Broy:2015zba}
B.~J.~Broy, D.~Ciupke, F.~G.~Pedro and A.~Westphal,
Starobinsky-Type Inflation from $\alpha'$-Corrections,
JCAP \textbf{01} (2016), 001
doi:10.1088/1475-7516/2016/01/001
[arXiv:1509.00024 [hep-th]].

\bibitem{Cicoli:2016xae}
M.~Cicoli, F.~Muia and P.~Shukla,
Global Embedding of Fibre Inflation Models,
JHEP \textbf{11} (2016), 182
doi:10.1007/JHEP11(2016)182
[arXiv:1611.04612 [hep-th]].

\bibitem{Cicoli:2017axo}
M.~Cicoli, D.~Ciupke, V.~A.~Diaz, V.~Guidetti, F.~Muia and P.~Shukla,
Chiral Global Embedding of Fibre Inflation Models,
JHEP \textbf{11} (2017), 207
doi:10.1007/JHEP11(2017)207
[arXiv:1709.01518 [hep-th]].

\bibitem{AbdusSalam:2022krp}
S.~AbdusSalam, C.~Crin{\`o} and P.~Shukla,
On K3-fibred LARGE Volume Scenario with de Sitter vacua from anti-D3-branes,
JHEP \textbf{03} (2023), 132
doi:10.1007/JHEP03(2023)132
[arXiv:2206.12889 [hep-th]].

\bibitem{Bera:2024ihl}
S.~Bera, D.~Chakraborty, G.~K.~Leontaris and P.~Shukla,
Global embedding of fiber inflation in a perturbative large volume scenario,
Phys. Rev. D \textbf{110} (2024) no.10, 106009
doi:10.1103/PhysRevD.110.106009
[arXiv:2406.01694 [hep-th]].

\bibitem{Broy:2014xwa}
B.~J.~Broy, F.~G.~Pedro and A.~Westphal,
Disentangling the $f(R)$ - Duality,
JCAP \textbf{03} (2015), 029
doi:10.1088/1475-7516/2015/03/029
[arXiv:1411.6010 [hep-th]].

\bibitem{Brinkmann:2023eph}
M.~Brinkmann, M.~Cicoli and P.~Zito,
Starobinsky inflation from string theory?,
JHEP \textbf{09} (2023), 038
doi:10.1007/JHEP09(2023)038
[arXiv:2305.05703 [hep-th]].

\bibitem{Berera:1995ie}
A.~Berera,
Warm inflation,
Phys. Rev. Lett. \textbf{75} (1995), 3218-3221
doi:10.1103/PhysRevLett.75.3218
[arXiv:astro-ph/9509049 [astro-ph]].

\bibitem{Kamali:2023lzq}
V.~Kamali, M.~Motaharfar and R.~O.~Ramos,
Recent Developments in Warm Inflation,
Universe \textbf{9} (2023) no.3, 124
doi:10.3390/universe9030124
[arXiv:2302.02827 [hep-ph]].

\bibitem{Berera:2023liv}
A.~Berera,
The Warm Inflation Story,
Universe \textbf{9} (2023) no.6, 272
doi:10.3390/universe9060272
[arXiv:2305.10879 [hep-ph]].

\bibitem{Das:2018rpg}
S.~Das,
Warm Inflation in the light of Swampland Criteria,
Phys. Rev. D \textbf{99} (2019) no.6, 063514
doi:10.1103/PhysRevD.99.063514
[arXiv:1810.05038 [hep-th]].

\bibitem{Motaharfar:2018zyb}
M.~Motaharfar, V.~Kamali and R.~O.~Ramos,
Warm inflation as a way out of the swampland,
Phys. Rev. D \textbf{99} (2019) no.6, 063513
doi:10.1103/PhysRevD.99.063513
[arXiv:1810.02816 [astro-ph.CO]].

\bibitem{Das:2019acf}
S.~Das, G.~Goswami and C.~Krishnan,
Swampland, axions, and minimal warm inflation,
Phys. Rev. D \textbf{101} (2020) no.10, 103529
doi:10.1103/PhysRevD.101.103529
[arXiv:1911.00323 [hep-th]].

\bibitem{Berera:2019zdd}
A.~Berera and J.~R.~Calder{\'o}n,
Trans-Planckian censorship and other swampland bothers addressed in warm inflation,
Phys. Rev. D \textbf{100} (2019) no.12, 123530
doi:10.1103/PhysRevD.100.123530
[arXiv:1910.10516 [hep-ph]].

\bibitem{Kamali:2019xnt}
V.~Kamali, M.~Motaharfar and R.~O.~Ramos,
Warm brane inflation with an exponential potential: a consistent realization away from the swampland,
Phys. Rev. D \textbf{101} (2020) no.2, 023535
doi:10.1103/PhysRevD.101.023535
[arXiv:1910.06796 [gr-qc]].

\bibitem{Berera:2020iyn}
A.~Berera, R.~Brandenberger, V.~Kamali and R.~Ramos,
Thermal, trapped and chromo-natural inflation in light of the swampland criteria and the trans-Planckian censorship conjecture,
Eur. Phys. J. C \textbf{81} (2021) no.5, 452
doi:10.1140/epjc/s10052-021-09240-3
[arXiv:2006.01902 [hep-th]].

\bibitem{Das:2020xmh}
S.~Das and R.~O.~Ramos,
Runaway potentials in warm inflation satisfying the swampland conjectures,
Phys. Rev. D \textbf{102} (2020) no.10, 103522
doi:10.1103/PhysRevD.102.103522
[arXiv:2007.15268 [hep-th]].

\bibitem{Brandenberger:2020oav}
R.~Brandenberger, V.~Kamali and R.~O.~Ramos,
Strengthening the de Sitter swampland conjecture in warm inflation,
JHEP \textbf{08} (2020), 127
doi:10.1007/JHEP08(2020)127
[arXiv:2002.04925 [hep-th]].

\bibitem{Motaharfar:2021egj}
M.~Motaharfar and R.~O.~Ramos,
Dirac-Born-Infeld warm inflation realization in the strong dissipation regime,
Phys. Rev. D \textbf{104} (2021) no.4, 043522
doi:10.1103/PhysRevD.104.043522
[arXiv:2105.01131 [hep-th]].

\bibitem{Bastero-Gil:2014raa}
M.~Bastero-Gil, A.~Berera, I.~G.~Moss and R.~O.~Ramos,
Theory of non-Gaussianity in warm inflation,
JCAP \textbf{12} (2014), 008
doi:10.1088/1475-7516/2014/12/008
[arXiv:1408.4391 [astro-ph.CO]].

\bibitem{Bartrum:2013fia}
S.~Bartrum, M.~Bastero-Gil, A.~Berera, R.~Cerezo, R.~O.~Ramos and J.~G.~Rosa,
The importance of being warm (during inflation),
Phys. Lett. B \textbf{732} (2014), 116-121
doi:10.1016/j.physletb.2014.03.029
[arXiv:1307.5868 [hep-ph]].

\bibitem{Benetti:2016jhf}
M.~Benetti and R.~O.~Ramos,
Warm inflation dissipative effects: predictions and constraints from the Planck data,
Phys. Rev. D \textbf{95} (2017) no.2, 023517
doi:10.1103/PhysRevD.95.023517
[arXiv:1610.08758 [astro-ph.CO]].

\bibitem{Bastero-Gil:2014oga}
M.~Bastero-Gil, A.~Berera, R.~O.~Ramos and J.~G.~Rosa,
Observational implications of mattergenesis during inflation,
JCAP \textbf{10} (2014), 053
doi:10.1088/1475-7516/2014/10/053
[arXiv:1404.4976 [astro-ph.CO]].

\bibitem{Berera:1998gx}
A.~Berera, M.~Gleiser and R.~O.~Ramos,
Strong dissipative behavior in quantum field theory,
Phys. Rev. D \textbf{58} (1998), 123508
doi:10.1103/PhysRevD.58.123508
[arXiv:hep-ph/9803394 [hep-ph]].

\bibitem{Yokoyama:1998ju}
J.~Yokoyama and A.~D.~Linde,
Is warm inflation possible?,
Phys. Rev. D \textbf{60} (1999), 083509
doi:10.1103/PhysRevD.60.083509
[arXiv:hep-ph/9809409 [hep-ph]].

\bibitem{Moss:2008yb}
I.~G.~Moss and C.~Xiong,
On the consistency of warm inflation,
JCAP \textbf{11} (2008), 023
doi:10.1088/1475-7516/2008/11/023
[arXiv:0808.0261 [astro-ph]].

\bibitem{delCampo:2010by}
S.~del Campo, R.~Herrera, D.~Pav{\'o}n and J.~R.~Villanueva,
On the consistency of warm inflation in the presence of viscosity,
JCAP \textbf{08} (2010), 002
doi:10.1088/1475-7516/2010/08/002
[arXiv:1007.0103 [astro-ph.CO]].

\bibitem{BasteroGil:2012zr}
M.~Bastero-Gil, A.~Berera, R.~Cerezo, R.~O.~Ramos and G.~S.~Vicente,
Stability analysis for the background equations for inflation with dissipation and in a viscous radiation bath,
JCAP \textbf{11} (2012), 042
doi:10.1088/1475-7516/2012/11/042
[arXiv:1209.0712 [astro-ph.CO]].

\bibitem{Berera:2008ar}
A.~Berera, I.~G.~Moss and R.~O.~Ramos,
Warm Inflation and its Microphysical Basis,
Rept. Prog. Phys. \textbf{72} (2009), 026901
doi:10.1088/0034-4885/72/2/026901
[arXiv:0808.1855 [hep-ph]].

\bibitem{BasteroGil:2012cm}
M.~Bastero-Gil, A.~Berera, R.~O.~Ramos and J.~G.~Rosa,
General dissipation coefficient in low-temperature warm inflation,
JCAP \textbf{01} (2013), 016
doi:10.1088/1475-7516/2013/01/016
[arXiv:1207.0445 [hep-ph]].

\bibitem{Bastero-Gil:2016qru}
M.~Bastero-Gil, A.~Berera, R.~O.~Ramos and J.~G.~Rosa,
Warm Little Inflaton,
Phys. Rev. Lett. \textbf{117} (2016) no.15, 151301
doi:10.1103/PhysRevLett.117.151301
[arXiv:1604.08838 [hep-ph]].

\bibitem{Bastero-Gil:2019gao}
M.~Bastero-Gil, A.~Berera, R.~O.~Ramos and J.~G.~Rosa,
Towards a reliable effective field theory of inflation,
Phys. Lett. B \textbf{813} (2021), 136055
doi:10.1016/j.physletb.2020.136055
[arXiv:1907.13410 [hep-ph]].

\bibitem{Berghaus:2019whh}
K.~V.~Berghaus, P.~W.~Graham and D.~E.~Kaplan,
Minimal Warm Inflation,
JCAP \textbf{03} (2020), 034
[erratum: JCAP \textbf{10} (2023), E02]
doi:10.1088/1475-7516/2020/03/034
[arXiv:1910.07525 [hep-ph]].

\bibitem{Laine:2021ego}
M.~Laine and S.~Procacci,
Minimal warm inflation with complete medium response,
JCAP \textbf{06} (2021), 031
doi:10.1088/1475-7516/2021/06/031
[arXiv:2102.09913 [hep-ph]].

\bibitem{Berghaus:2025dqi}
K.~V.~Berghaus, M.~Drewes and S.~Zell,
Warm Inflation with the Standard Model,
[arXiv:2503.18829 [hep-ph]].

\bibitem{Cicoli:2018cgu}
M.~Cicoli and G.~A.~Piovano,
Reheating and Dark Radiation after Fibre Inflation,
JCAP \textbf{02} (2019), 048
doi:10.1088/1475-7516/2019/02/048
[arXiv:1809.01159 [hep-th]].

\bibitem{Cicoli:2021rub}
M.~Cicoli, F.~Quevedo, R.~Savelli, A.~Schachner and R.~Valandro,
Systematics of the {\ensuremath{\alpha}}' expansion in F-theory,
JHEP \textbf{08} (2021), 099
doi:10.1007/JHEP08(2021)099
[arXiv:2106.04592 [hep-th]].

\bibitem{AbdusSalam:2020ywo}
S.~AbdusSalam, S.~Abel, M.~Cicoli, F.~Quevedo and P.~Shukla,
A systematic approach to K{\"a}hler moduli stabilisation,
JHEP \textbf{08} (2020) no.08, 047
doi:10.1007/JHEP08(2020)047
[arXiv:2005.11329 [hep-th]].

\bibitem{Grimm:2017okk}
T.~W.~Grimm, K.~Mayer and M.~Weissenbacher,
Higher derivatives in Type II and M-theory on Calabi-Yau threefolds,
JHEP \textbf{02} (2018), 127
doi:10.1007/JHEP02(2018)127
[arXiv:1702.08404 [hep-th]].

\bibitem{Ramos:2013nsa}
R.~O.~Ramos and L.~A.~da Silva,
Power spectrum for inflation models with quantum and thermal noises,
JCAP \textbf{03} (2013), 032
doi:10.1088/1475-7516/2013/03/032
[arXiv:1302.3544 [astro-ph.CO]].

\bibitem{Montefalcone:2023pvh}
G.~Montefalcone, V.~Aragam, L.~Visinelli and K.~Freese,
WarmSPy: a numerical study of cosmological perturbations in warm inflation,
JCAP \textbf{01} (2024), 032
doi:10.1088/1475-7516/2024/01/032
[arXiv:2306.16190 [astro-ph.CO]].

\bibitem{Rodrigues:2025neh}
G.~S.~Rodrigues and R.~O.~Ramos,
WI2easy: warm inflation dynamics made easy,
[arXiv:2504.17760 [astro-ph.CO]].

\bibitem{BasteroGil:2010pb}
M.~Bastero-Gil, A.~Berera and R.~O.~Ramos,
Dissipation coefficients from scalar and fermion quantum field interactions,
JCAP \textbf{09} (2011), 033
doi:10.1088/1475-7516/2011/09/033
[arXiv:1008.1929 [hep-ph]].

\bibitem{ORamos:2025uqs}
R.~O.~Ramos and G.~S.~Rodrigues,
Viability of warm inflation with standard model interactions,
Phys. Rev. D \textbf{111} (2025) no.12, 123527
doi:10.1103/wn1m-19gt
[arXiv:2504.20943 [hep-ph]].

\bibitem{Liddle:2003as}
A.~R.~Liddle and S.~M.~Leach,
How long before the end of inflation were observable perturbations produced?,
Phys. Rev. D \textbf{68} (2003), 103503
doi:10.1103/PhysRevD.68.103503
[arXiv:astro-ph/0305263 [astro-ph]].

\bibitem{Planck:2018vyg}
N.~Aghanim \textit{et al.} [Planck],
Planck 2018 results. VI. Cosmological parameters,
Astron. Astrophys. \textbf{641} (2020), A6
[erratum: Astron. Astrophys. \textbf{652} (2021), C4]
doi:10.1051/0004-6361/201833910
[arXiv:1807.06209 [astro-ph.CO]].

\bibitem{Ballesteros:2022hjk}
G.~Ballesteros, M.~A.~G.~Garc\'\i{}a, A.~P.~Rodr\'\i{}guez, M.~Pierre and J.~Rey,
Primordial black holes and gravitational waves from dissipation during inflation,
JCAP \textbf{12} (2022), 006
doi:10.1088/1475-7516/2022/12/006
[arXiv:2208.14978 [astro-ph.CO]].

\bibitem{Ballesteros:2023dno}
G.~Ballesteros, A.~Perez Rodriguez and M.~Pierre,
Monomial warm inflation revisited,
JCAP \textbf{03} (2024), 003
doi:10.1088/1475-7516/2024/03/003
[arXiv:2304.05978 [astro-ph.CO]].

\bibitem{BICEP:2021xfz}
P.~A.~R.~Ade \textit{et al.} [BICEP and Keck],
Improved Constraints on Primordial Gravitational Waves using Planck, WMAP, and BICEP/Keck Observations through the 2018 Observing Season,
Phys. Rev. Lett. \textbf{127} (2021) no.15, 151301
doi:10.1103/PhysRevLett.127.151301
[arXiv:2110.00483 [astro-ph.CO]].

\bibitem{ACT:2025tim}
E.~Calabrese \textit{et al.} [ACT],
The Atacama Cosmology Telescope: DR6 Constraints on Extended Cosmological Models,
[arXiv:2503.14454 [astro-ph.CO]].

\bibitem{ACT:2025fju}
T.~Louis \textit{et al.} [ACT],
The Atacama Cosmology Telescope: DR6 Power Spectra, Likelihoods and $\Lambda$CDM Parameters,
[arXiv:2503.14452 [astro-ph.CO]].

\bibitem{Berera:2025vsu}
A.~Berera, S.~Brahma, Z.~Qiu, R.~O.~Ramos and G.~S.~Rodrigues,
The early universe is $\textit{ACT}$-ing $\textit{warm}$,
[arXiv:2504.02655 [hep-th]].

\bibitem{Das:2022ubr}
S.~Das and R.~O.~Ramos,
Running and Running of the Running of the Scalar Spectral Index in Warm Inflation,
Universe \textbf{9} (2023) no.2, 76
doi:10.3390/universe9020076
[arXiv:2212.13914 [astro-ph.CO]].

\bibitem{Planck:2018jri}
Y.~Akrami \textit{et al.} [Planck],
Planck 2018 results. X. Constraints on inflation,
Astron. Astrophys. \textbf{641} (2020), A10
doi:10.1051/0004-6361/201833887
[arXiv:1807.06211 [astro-ph.CO]].

\bibitem{Obied:2018sgi}
G.~Obied, H.~Ooguri, L.~Spodyneiko and C.~Vafa,
De Sitter Space and the Swampland,
[arXiv:1806.08362 [hep-th]].

\bibitem{Agrawal:2018own}
P.~Agrawal, G.~Obied, P.~J.~Steinhardt and C.~Vafa,
On the Cosmological Implications of the String Swampland,
Phys. Lett. B \textbf{784} (2018), 271-276
doi:10.1016/j.physletb.2018.07.040
[arXiv:1806.09718 [hep-th]].

\bibitem{Garg:2018reu}
S.~K.~Garg and C.~Krishnan,
Bounds on Slow Roll and the de Sitter Swampland,
JHEP \textbf{11} (2019), 075
doi:10.1007/JHEP11(2019)075
[arXiv:1807.05193 [hep-th]].

\bibitem{Palti:2019pca}
E.~Palti,
The Swampland: Introduction and Review,
Fortsch. Phys. \textbf{67} (2019) no.6, 1900037
doi:10.1002/prop.201900037
[arXiv:1903.06239 [hep-th]].

\bibitem{Berera:1999ws}
A.~Berera,
Warm inflation at arbitrary adiabaticity: A Model, an existence proof for inflationary dynamics in quantum field theory,
Nucl. Phys. B \textbf{585} (2000), 666-714
doi:10.1016/S0550-3213(00)00411-9
[arXiv:hep-ph/9904409 [hep-ph]].

\bibitem{Berera:2003yyp}
A.~Berera,
Warm inflation solution to the eta problem,
PoS \textbf{AHEP2003} (2003), 069
doi:10.22323/1.010.0069
[arXiv:hep-ph/0401139 [hep-ph]].

\bibitem{Ito:2025lcg}
A.~Ito and R.~O.~Ramos,
Warm multi natural inflation,
[arXiv:2504.15606 [hep-ph]].

\end{thebibliography}
\end{document}